\documentclass[preprint,12pt]{elsarticle}

\usepackage{amssymb}
\usepackage{multirow}
\usepackage{float}
\usepackage{caption, subcaption}
\usepackage{amsmath}
\usepackage{mathtools}
\usepackage{physics}
\usepackage{xcolor}
\usepackage[utf8]{inputenc}

\begin{document}

\begin{frontmatter}



\title{Critical behaviour near  critical end points and  tricritical points in disordered spin-1 ferromagnets }


\author[inst1,inst2,inst3]{Soheli Mukherjee}

\affiliation[inst1]{ organization={Department of Solar Energy and Environmental Physics},  addressline={ Blaustein Institutes for Desert Research, Ben-Gurion University of
  the Negev}, city= {Sede Boqer Campus},  postcode={8499000}, country={Israel}}

\affiliation[inst2]{organization={School of Physical Sciences},
            addressline={National Institute of Science Education and Research, Bhubaneswar}, 
            city={P.O. Jatni},
            postcode={752050}, 
            state={Odisha},
            country={India}}

\author[inst2,inst3]{Sumedha}

\affiliation[inst3]{organization={Homi Bhabha National Institute},
            addressline={Training School Complex}, 
            city={Anushaktinagar},
            postcode={400094}, 
            state={Mumbai},
            country={India}}

\begin{abstract}
 Critical end points and tricritical points are multicritical points that separate  lines of continuous transitions from  lines of first order transitions in the phase diagram of many systems. In models like the spin-1 disordered Blume-Capel model and  the repulsive Blume-Emery-Griffiths model, the tricritical point splits into a critical end point and a bicritical end point with an increase in disorder and repulsive coupling strength respectively. In order to make a distinction between these two multicritical points, we investigate and contrast the  behaviour of the first order phase boundary and the co-existence diameter around them.  
\end{abstract}




 \end{frontmatter}


\section{Introduction} \label{sec:sample1}

In many physical systems, the order of the  phase transition changes with the change of couplings in the system. In systems like  alloys of magnetic and non-magnetic materials \cite{PhysRevB.80.014427},  superconducting films \cite{PhysRevLett.30.1038}, quantum metals \cite{PhysRevLett.94.247205}, polymer collapse \cite{de1975collapse}, polymerized membranes \cite{WIESE1995495}, liquid crystal \cite{shashidhar1975pressure}, metamagnets \cite{PhysRevB.1.2250, keen1966first, PhysRev.164.866},   piezoelectric materials \cite{PhysRevLett.103.257602} a line of continuous transitions changes to a line of first order transitions via a multicritical point which can either be a tricritical point (TCP) or a critical end point (CEP).  TCPs are the most widely studied and understood multicritical point.  TCP is a point where three phases become identical.  Alternatively, it can also be defined as a multicritical point at which a line of continuous transitions ends and a line of first order transitions originates.

CEP, on the other hand, is a critical point where a line of second order transitions abruptly  terminates at a line of first order transitions \cite{widom1973tricritical, chaikin_lubensky_1995}. Alternately, it is a point where two phases become critical in the presence of one or more ordered phases, known as the spectator phases \cite{PhysRevLett.65.2402}, in systems with multiple phases. There are a variety of physical systems whose phase diagrams contain CEP. For example, superfluids \cite{PhysRevResearch.2.013194}, gel-fluid mixture \cite{PhysRevLett.40.820}, metamagnets \cite{doi:10.1143/JPSJ.80.093707}, ferroelectrics \cite{Iwata_2021}, liquid crystals \cite{doi:10.7566/JPSJ.91.024002}, binary fluid mixtures \cite{PhysRevLett.67.1555, PhysRevLett.78.1488, PhysRevE.55.6624}, quantum chromodynamics \cite{AYALA201577}. There are systems as well, for example, free standing smectic films \cite{PhysRevE.84.061706}, spin crossover materials \cite{PhysRevB.93.064109} where crossover happens from a TCP to a CEP.

Both the TCP and CEP separate a line of critical points from a line of first order transitions which makes it tricky to distinguish them in experiments and simulations. A TCP is where a line of continuous transitions meets at the peak of a co-existence region. On the contrary, a CEP is a point where a line of second order transitions truncates on the either side of the peak of the  co-existence region. The scaling and universality class near the TCPs are well understood and have been verified in many models \cite{PhysRevLett.28.675, PhysRevLett.29.349, PhysRevLett.29.278, domb1984phase}. TCP lies in a different universality class than the critical point. The sixth order Landau theory captures the behaviour near a TCP completely at the mean-field level. On the other hand, the bulk critical exponents near a CEP are 
 expected to be the same as that of a critical point \cite{doi:10.1063/1.1680768, DIEHL2000268, diehl2001bulk}. There are though further singularities that arise near a CEP in the bulk \cite{ fisher1990phases, PhysRevB.43.11177, PhysRevB.43.10635, PhysRevB.45.5199, FISHER199177} and  surface \cite{PhysRevLett.65.2402, PhysRevLett.65.3405, doi:10.1080/00268970500151338, PhysRevE.71.011601}.    Using scaling arguments Fisher showed that the non-analyticities of the first order surface near a CEP are related to the singularities of the critical line \cite{fisher1990phases}. This prediction was verified using spherical models \cite{PhysRevB.43.11177, PhysRevB.43.10635, PhysRevB.45.5199, FISHER199177}, using eighth order Landau free energy expansion for isomorphous transitions \cite{ishibashi1991isomorphous, DESANTAHELENA1994479} and two order parameter Landau free energy expansion for ferroelectric materials \cite{DESANTAHELENA1995408}. Later, using extensive Monte-Carlo simulations Wilding provided the first evidence of the singular behaviour on the first order phase boundary near a CEP for a symmetrical binary fluid \cite{PhysRevLett.78.1488, PhysRevE.55.6624}. Both the Fisher and the Wilding conjecture were verified for an asymmetric Ising  model using Wang-Landau study on a triangular lattice \cite{PhysRevE.75.061108, landau2008critical}.

The analysis of the above-mentioned scaling relations has been done only on pure models. In disordered spin  systems with TCP in the pure case, CEP can emerge as the system gets more disordered.  In such cases the CEP arises due to the splitting of TCP into a CEP and a bicritical end point  (BEP) \cite{mukherjee2020emergence, mukherjee2022phase}. It hence becomes important to find observables that can distinguish TCP and CEP in simulations and experiments. We have studied the phase diagrams of the spin-1 Blume-Capel model in the presence of quenched  disorder \cite{mukherjee2020emergence} and in the presence of repulsive biquadratic exchange interactions  \cite{Mukherjee_2021}. Both of these models exhibit TCP as well as CEP depending on the strength of the disorder or higher order interactions. We study the behaviour of the phase boundary and the change in the co-existence curve  for these models in detail in this paper. We test the Fisher and Wilding's scaling hypothesis near CEP and also show how these can be used to distinguish CEP and TCP.

 The article is organized  as follows.  In section 2, we introduce the spin-1 models we used for the study and briefly review the phase diagrams of the models. In section 3, we discuss the singularities in the phase boundary  near a CEP.  We study the singularities of the phase boundaries near a CEP for the spin-1 models and compare them with the behaviour  of similar quantities   near a TCP.  In section 4, we  discuss the singularities in the co-existence curve and
 study the co-existence curve near a CEP for the spin-1 models and compare the corresponding quantities near a TCP. We then summarize the conclusions briefly in section 5.

\section{Model}\label{sec2}

We study two models: random crystal field Blume Capel (RCFBC) model \cite{mukherjee2020emergence} and repulsive Blume-Emery-Griffiths (RBEG)  model  \cite{Mukherjee_2021}.  We briefly discuss the phase diagram of these two models in this section.

\subsection{Random crystal field Blume-Capel model (RCFBC)}

The Hamiltonian of the system on a fully connected graph is \cite{mukherjee2020emergence}
\begin{equation}\label{eq510}
\mathcal{H} (C_N)= -\frac{1}{2 N} (\sum_i s_i)^2  +  \sum_i \Delta_i s_i ^2 - H \sum_i  s_i
\end{equation}
where  $H$ is the external magnetic field, $s_i$ are the spin$-1$ random variables which can take $\pm 1,0$ values and $\Delta_i$ represents the quenched random crystal field at the i$^{th}$ site and is an i.i.d from a bimodal distribution
\begin{equation}
P(\Delta_i) = (1-p) \,\,  \delta(\Delta - \Delta_i) + p \,\,  \delta(\Delta + \Delta_i)
\end{equation}
 where $0 \leq p \leq 1$. The order parameters of the system are $m = \langle s \rangle$ and $q = \langle s^2 \rangle$. The model was solved in \cite{mukherjee2020emergence, Sumedha_2016} using  large deviations theory (LDT). The free energy functional of the model is given by
\begin{eqnarray} \label{eq26}
	f(m) =  \frac{\beta m^2}{2} &-& (1-p) \log (1+2 e^{-\beta \Delta} \cosh { \beta (m + H)}  ) \nonumber \\
		& -&  p \log (1+2 e^{\beta \Delta} \cosh { \beta (m + H)}  )
	\end{eqnarray}
where $\beta = \frac{1}{T}$ is the inverse temperature. Minima of $f(m)$ over $m$ gives the free energy of the system. The equation of minima satisfies the following self-consistent equation
\begin{equation}\label{eqmag2}
    m = 2 \sinh{\beta (m+H)} \Bigg ( \frac{ p \,\,  e^{\beta \Delta}}{1 + 2 e^{\beta \Delta} \cosh{\beta (m+H)}} + \frac{(1-p) \,\,  e^{- \beta \Delta}}{1 + 2 e^{- \beta \Delta} \cosh{\beta (m+H)}} \Bigg ) 
\end{equation}
At the fixed points the $m$ and $q$ are connected via the following equation
\begin{equation}\label{eqmag}
    \tanh(\beta (m+H))=\frac{m}{q}
\end{equation}

The equation of the continuous transition line (also known as the $\lambda$ line)  in the $T-\Delta$ plane can be obtained via linear stability analysis and comes out to be \cite{mukherjee2020emergence, Sumedha_2016}

\begin{equation}\label{lline}
5-4 \beta = 2 (\beta p-1) e^{\beta \Delta}+2(\beta-\beta p-1) e^{-\beta \Delta}
\end{equation}

In the $T-q$ plane, the equation has a rather simple form
\begin{equation}\label{eqq1}
    q = \frac{1}{\beta}
\end{equation}

\begin{figure}
     \centering
     \begin{subfigure}[b]{0.495\textwidth}
         \centering
         \includegraphics[width= \textwidth]{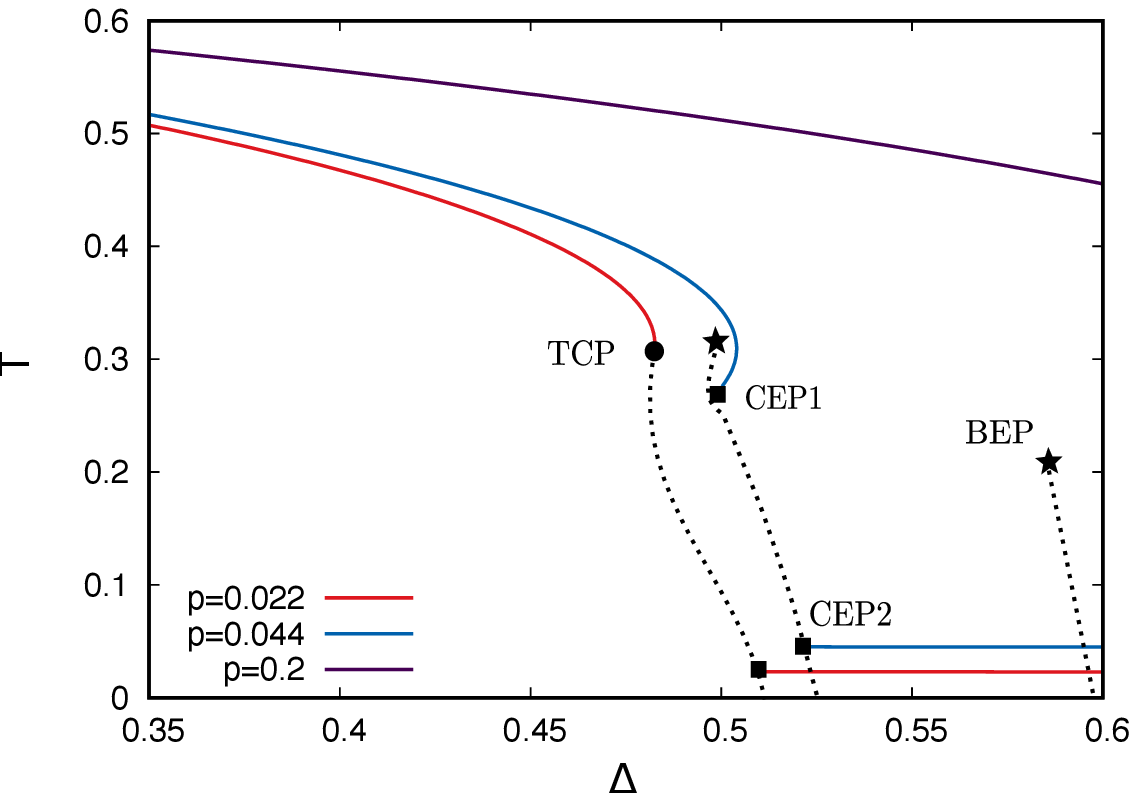}
         \caption{ }
        \label{fig620new}
     \end{subfigure}
     \hfill
     \begin{subfigure}[b]{0.495\textwidth}
         \centering
        \includegraphics[width= \textwidth]{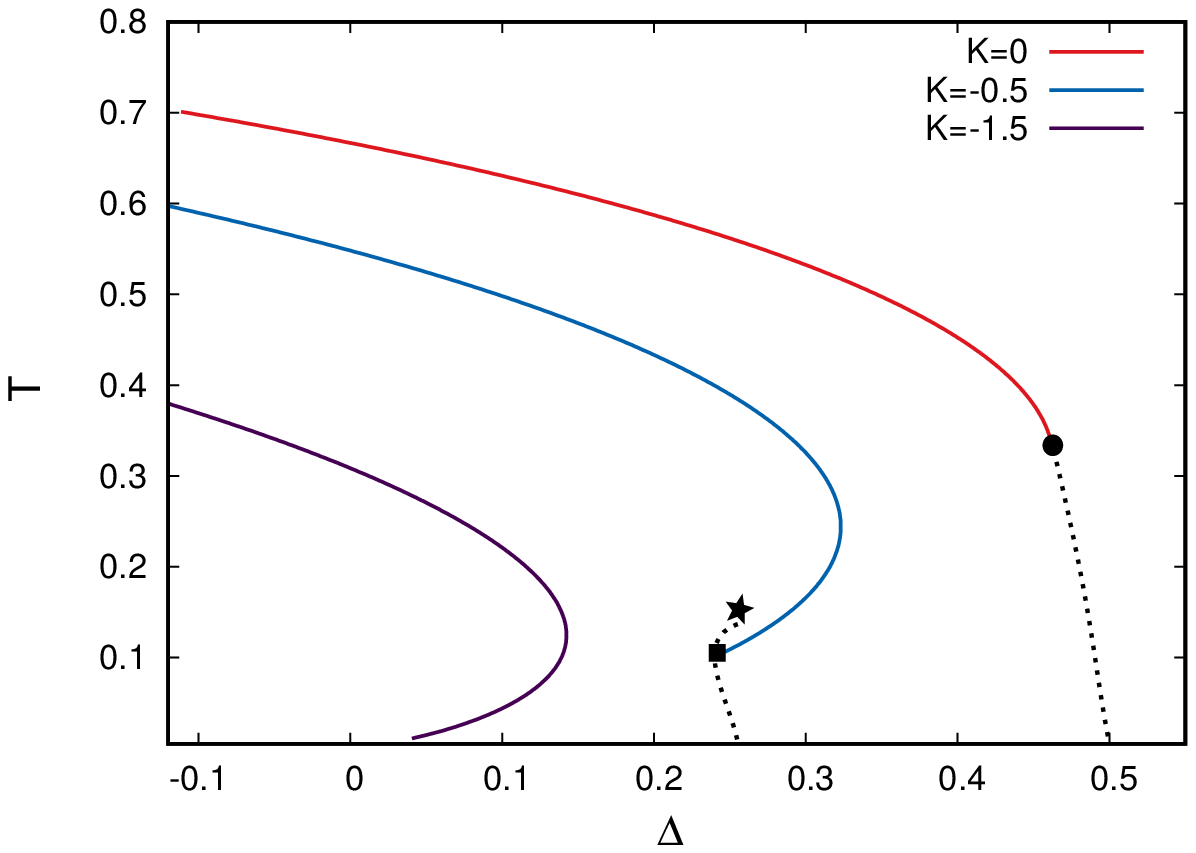}
         \caption{}
         \label{fig61new}
    \end{subfigure}
  \hfill
    \caption{ (colour online)  The three different phase diagrams of the RCFBC and the RBEG model in the $T$ - $\Delta$  plane ($H=0$ plane). The solid line denotes the lines of continuous transitions, the dotted line denotes the lines of first order transitions, the solid circles are the TCPs, solid squares are the CEPs, and the solid stars are the BEPs.  \textbf{(a)} shows the phase diagrams for the RCFBC model for three ranges of $p$: the weak, intermediate and strong disorder regimes. \textbf{(b)} shows the phase diagrams of the RBEG model for three different values of $K$.}
      \label{fig61newnn}
\end{figure}

The phase diagram of the RCFBC was classified  into three categories in \cite{mukherjee2020emergence} depending on the strength of disorder $p$: weak ($0 < p \leq  0.022$), intermediate ($0.022 < p \leq  0.107875$ ), and strong ($0.107875 < p \leq  0.5$). These are   shown in Fig.\ref{fig620new} in the $T-\Delta$ plane. For weak  strength of the disorder, the phase diagram consists of a line of first order transitions, denoted by the dotted line and two lines of  second order transitions, denoted by the solid red line for $p=0.022$.  The lines of first order and second order  transitions meet at a TCP at higher temperature and CEP at a lower temperature (we call it CEP2).  For intermediate strength of disorder, the TCP  branches into a CEP (we call this CEP1) and a BEP  and the phase diagram exhibits two CEPs and one BEP (shown for $p=0.044$).  As disorder increases further, in the strong disorder regime the CEPs vanish and the phase diagram consists of a BEP inside the ordered regime (shown for $p=0.2$).


\subsection{Repulsive Blume-Emery-Griffiths model (RBEG)}


The Hamiltonian of the RBEG model on a fully connected graph is  \cite{Mukherjee_2021}

\begin{equation}\label{eq5}
\mathcal{H}= -\frac{1}{2 N} (\sum_i s_i)^2 - \frac{K}{2 N} (\sum_i s_i^2)^2  + \Delta \sum_i s_i ^2 - H \sum_i  s_i
\end{equation}
here $K$ is the biquadratic exchange interaction term.  This model was solved in \cite{Mukherjee_2021, PhysRevE.100.052135}.  The free energy functional of the RBEG model at the fixed points is
\begin{eqnarray}
f(m) &=& \frac{\beta m^{2}}{2} +\frac{\beta K m^2 \coth^2{\beta m}}{2} +\log (1+2 e^{-\beta \Delta} \cosh \beta H) \nonumber \\
&-& \log (1+2 e^{\beta(K m \coth{\beta m}-\Delta)} \cosh \beta (m+H) ) 
\label{freeenergy}
\end{eqnarray}
Similar to the RCFBC model, one of the fixed point equations of the RBEG model is given by Eq. \ref{eqmag}. The other one is 
\begin{equation}
    2 (\beta -1) = e^{\beta \Delta - K}
\end{equation}
As a result the $\lambda$ line equation in the $T-q$ plane is again $q = \frac{1}{\beta}$ (same as Eq. \ref{eqq1}).

In the presence of repulsive biquadratic exchange interaction ($K < 0$), the phase diagram shows similar behaviour like the RCFBC model in the weak and intermediate disorder regime for ``$ 0 < K \leq - 0. 1838$" and ``$ - 0. 1838 < K \leq - 1$" respectively. The phase diagrams are shown in Fig. \ref{fig61new} for $K=0$ and $K=-0.5$, respectively. At exactly $K = -1$, the first order transition line as well as the multicritical points move to $T=0$ and for all $K < -1$, the phase diagram consists only of a second order line of transitions, shown in Fig. \ref{fig61new} (for $K=-1.5$).

\begin{figure}
     \centering
     \begin{subfigure}[b]{0.32\textwidth}
         \centering
         \includegraphics[width= \textwidth]{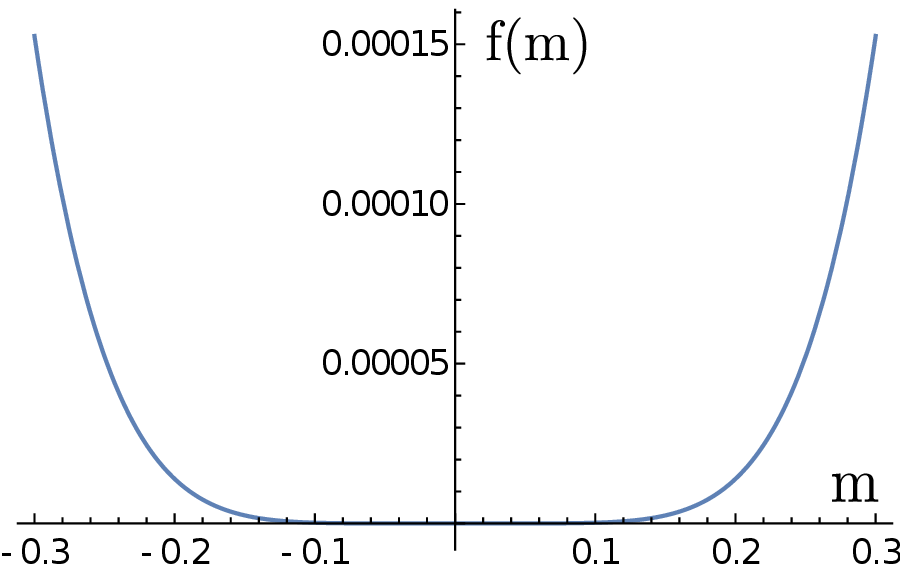}
         \caption{ }
        \label{fig69new1}
     \end{subfigure}
     \hfill
     \begin{subfigure}[b]{0.32\textwidth}
         \centering
        \includegraphics[width= \textwidth]{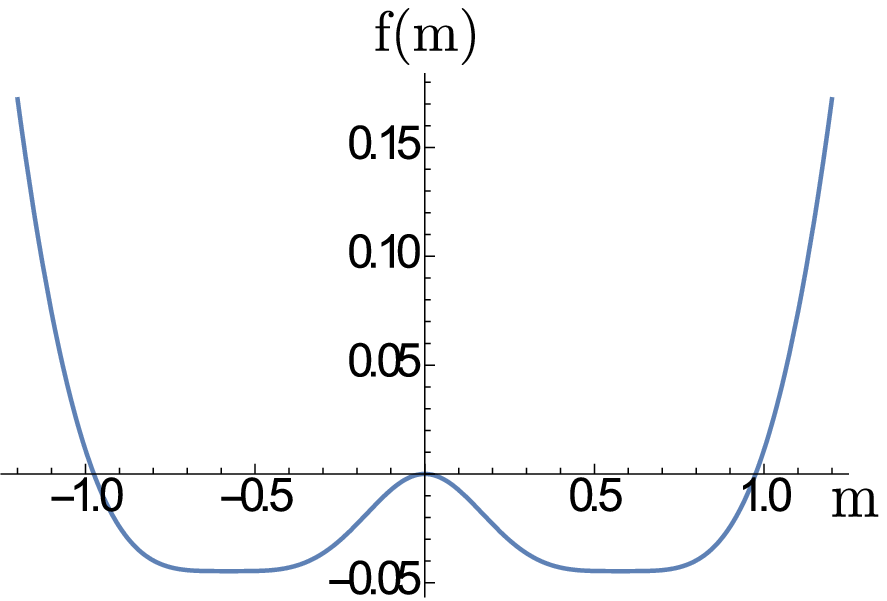}
         \caption{}
         \label{fig69new2}
    \end{subfigure}
  \hfill
  \begin{subfigure}[b]{0.32\textwidth}
         \centering
        \includegraphics[width= \textwidth]{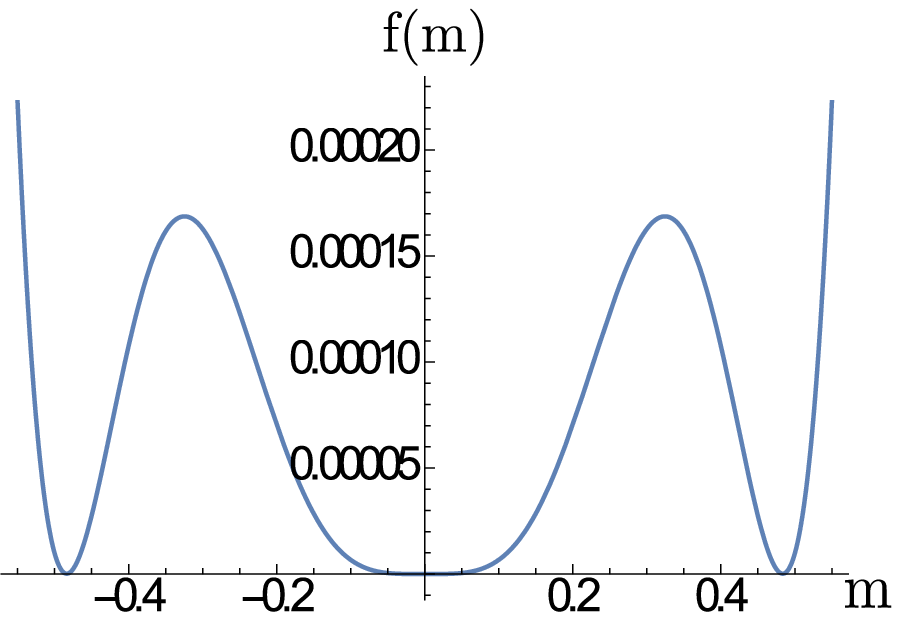}
         \caption{}
         \label{fig69new3}
    \end{subfigure}
  \hfill
    \caption{ The free energy functional $f(m)$ (given by Eq. \ref{eq26}) plotted as a function of  $m$ at the \textbf{(a)} TCP, \textbf{(b)} BEP and \textbf{(c)} CEP.}
      \label{fig69new}
\end{figure}

Fig. \ref{fig69new} shows the plot of the free energy functional $f(m)$ as a function of $m$ at the multicritical points. The  TCP is a point where three phases coalesce into one phase, as shown in Fig. \ref{fig69new1}. Using a $6^{th}$ order Landau free energy  expansion as long as the $6^{th}$ order coefficient is positive, the TCP of a system can be located correctly. On the other hand, the BEP is a point of co-existence of two critical phases and CEP is a point of co-existence of a critical phase with one or more non-critical phases, shown in Fig. \ref{fig69new2} and Fig. \ref{fig69new3} respectively. Landau free energy expansion upto $8^{th}$ order allows for the existence of CEPs and BEPs. The Landau free energy expansion of the $f(m)$ till $8^{th}$ order is given by

\begin{equation} \label{eq:16}
f(m)=a_2 m^2 + a_4 m^4+ a_6 m^6+ a_8 m^8
\end{equation}
where $a_i$'s are the Landau coefficients. According to the Landau theory, the TCP occurs when $a_2=a_4=0$ with the condition $a_6>0$. The conditions for BEP and CEP  for $8^{th}$ order Landau theory are \cite{ ishibashi1991isomorphous, mukherjee2020emergence}

\[
a_6 <0,  a_8 > 0 \,\,\, \text{and}, 
\begin{cases}
  \frac{a_6^2}{4 a_4 a_8} = 1 & \text{For CEP} \\
  \frac{3 a_6^2}{8 a_4 a_8} = 1 & \text{For BEP}
\end{cases}
\]
Since both the CEP and the BEP are points of co-existence, the truncation of the Landau free energy expansion does not locate them correctly \cite{mukherjee2020emergence}.   Using the full free energy function $f(m)$, we showed that  BEP can be located correctly using the condition $f'(m)=f''(m)=f'''(m)=0$ and $f''''(m)>0$, and the CEP can be located correctly  where the first order line measured by equating $f(m=0)=f(m \neq 0)$ and $f'(m=0)=f'(m \neq 0)$ satisfies the $\lambda$ line equation (Eq. \ref{lline}) \cite{mukherjee2020emergence}.

\section{Study of the singularity in the phase boundary near a CEP}

Near a critical point (CP) the measurable thermodynamic quantities such as the $m$, magnetic susceptibility $\chi$, specific heat $C_v$ etc, show a power law behaviour. The singularities near a CP (at $T = T_c$) are expressed in terms of the universal critical exponents and critical amplitudes as

\begin{eqnarray}
   m \approx D \,\, |t|^{\beta} \label{eq:6}\\
   m \approx B_{\pm} \,\, H^{1/\delta}  \label{eq:7}\\ 
    \chi \approx  C_{\pm} \,\, |t|^{- \gamma} \label{eq:8} \\
     C_v \approx  A_{\pm} \,\, |t|^{- \alpha} \label{eq:9}
\end{eqnarray}
here  $t=\frac{T-T_c}{T_c}$ is the scaled temperature and $H$ is the ordering field (for example, the external magnetic field in case of a ferromagnet). $\alpha, \beta, \gamma, \delta$ are the critical exponents  of the system. They are related to each other by the relations
\begin{eqnarray}
   \alpha + 2 \beta + \gamma =2  \label{eq:10}\\
  \beta \gamma = \beta + \gamma
\end{eqnarray}
known as the Rushbrooke and the Widom identity respectively \cite{goldenfeld2018lectures}. The coefficients $A_\pm$, $B_\pm$, $C_\pm$ are the critical amplitudes. They are non-universal but their ratios are  universal. For example, for  Ising model $\frac{C_+}{C_-} = 2$ (mean-field),    $\frac{C_+}{C_-}  \approx 4.95$ and  $\frac{A_+}{A_-} \approx 0.523$ (in three dimension) \cite{LIU198935, PhysRevB.11.1217}.

The Ising TCP universality class is different than the Ising universality class \cite{domb1984phase}. The mean-field values of the critical exponents near a TCP are $\beta=0.25$, $\gamma= 1$ and $\delta=5$. Whereas near a CP these are $\beta = \frac{1}{2}$, $ \gamma= 1$ and $\delta =3$.

Contrary to a TCP, a CEP is a point where the $\lambda$ line abruptly truncates at the co-existence curve.  The critical exponents near a CEP thus fall under the same universality class as a CP. The CEP though differs from a  CP, as near a CEP, the curvature of the phase boundary exhibits a singularity that can be expressed in terms of the universal amplitude ratios of the critical behaviour of the critical line ($\lambda$ line) \cite{ fisher1990phases, PhysRevB.43.11177, PhysRevB.43.10635, PhysRevB.45.5199, FISHER199177}.  In the next  subsections, we briefly review these scaling  by Fisher \textit{et al} \cite{fisher1990phases, PhysRevB.43.11177}  near a CEP. We then verify them near CEP in RCFBC and RBEG model. We also study the same quantities near a TCP.

\begin{figure}
\centering
\includegraphics[scale=0.5]{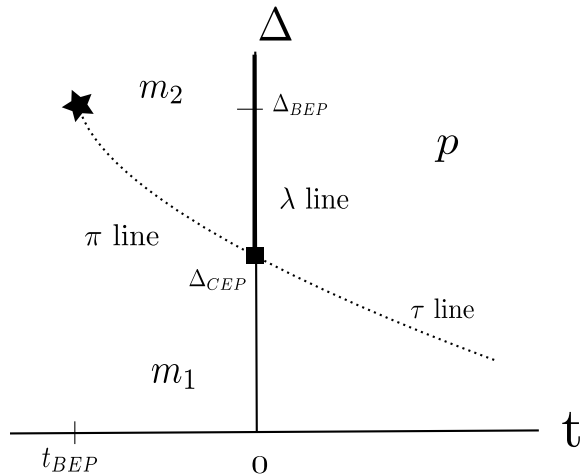}
\caption{Schematic phase diagram of the spin-1 ferromagnets in the $\Delta-t$ plane, where  $t=\frac{T-T_c(\Delta)}{T_c(\Delta)}$  is the the scaled temperature. The solid line depicts lines of second order transition and dotted lines are the lines of first order transition. The first order phase boundary $\pi$ separates the two ordered phases $m_1$ and $m_2$ and the $\tau$ separates  the phase $m_1$ and the disordered phase $p$. The solid star represents BEP and the solid square is CEP. Figure adapted from \cite{DESANTAHELENA1994479}.  }
\label{fig610}
\end{figure}

\subsection{Singularities of the phase boundary near a CEP}

As the $\lambda$ line is approached in the $T-\Delta-H$ space (here $\Delta$ is the non-ordering field of the system. For example, in a binary fluid mixture it is the chemical potential, in liquid crystals it is a geometrical parameter, and for spin-1 systems it is the crystal field), the free energy can be expressed as the sum of an analytic boundary term and a singular term \cite{PhysRevB.43.11177}

\begin{equation}\label{criticaleq}
 f(T, \Delta, H) \sim f_{ns}(T, \Delta, H)- Q \,\, |t|^{2-\alpha} \,\,  W_\pm \Bigg [\frac{ U H}{|t|^{\Delta}} \Bigg ]
 \end{equation}
where $f_{ns}(T,\Delta, H)$ is the non-singular part of the free energy,  $+(-)$ sign refers to $T>T_c$ ($T<T_c$). $W_\pm$ is the scaling function, and $Q$ and $U$ are the analytic functions of $T$, $\Delta$ and $H$. 

 Fig. \ref{fig610} shows the schematic plot of a phase diagram exhibiting a CEP in the  $\Delta$ - $t$  plane.  The $\pi$ line is the first order transition line along which two ferromagnetic states $m_1$ and $m_2$ co-exist. Similarly, the $\tau$ line is the line of co-existence of a disordered phase ($p$) and an ordered phase ($m_1$). The $p$ phase and the $m_2$ phase become critical in the presence of a non-critical phase $m_1$ at the CEP. $m_1$ is called the spectator phase. The separation between the two ordered phases ($m_1$ and $ m_2$) becomes zero at the BEP. This is  shown by a black star in Fig. \ref{fig610}.

In presence of an ordering field $H$, the first order phase boundary becomes a surface $\rho$ between the critical and noncritical phases in the  $t-\Delta-H$ space, specified by the function $\Delta_{\rho} (T, H)$. It was shown by Fisher  \textit{et.al} \cite{fisher1990phases, PhysRevB.43.11177} that the curvature of the first order surface $\rho$ shows singularity and the equation of the surface $\rho$ can be derived by equating the free energies of the phases near a  CEP. So in the vicinity of a CEP ($T_{CEP}, \,\, \Delta_{CEP}, 0$) the equation of the phase boundary $\rho$ is expressed as

\begin{equation}\label{eq-fisher}
 \Delta_\rho(T,H) =  \Delta_{CEP} + \Delta_1 \,\, \hat t + \Delta_2 \,\, H -X_\pm  \,\,|\hat t |^{2-\alpha} - Y \,\, |H| \,\,  |\hat t|^ \beta - \frac{1}{2} Z_\pm \,\,  H^2 \,\,  |\hat t|^{- \gamma} + ...
\end{equation}
here $\hat t = \frac{T- T_{CEP}}{T_{CEP}}$ and $\Delta_1$, \, $\Delta_2$ are the non-singular functions of $\hat{t}$ and $H$. The amplitudes $X_\pm$, $Y$, $Z_\pm$ are related to the curvature of the phase boundary as follows

\begin{eqnarray}
 X_\pm \approx |e_0|^{2 - \alpha} W_\pm(0) \\
 Z_\pm \approx  |e_0|^{\gamma} W_\pm''(0) \\
 Y \approx  |e_0|^{\beta} W_{-}'(0) 
\end{eqnarray}
$W_\pm$ is the scaling function of the free energy given (Eq. \ref{criticaleq}) and $e_0$ is the geometrical factor defined as
\begin{equation}
    e_0 =  1- \Bigg [ \pdv{\Delta_\rho}{T} \Bigg]_{CEP}  \,\, \Bigg [ \pdv{T_c(\Delta)}{\Delta} \Bigg]_{CEP} 
\end{equation}

Hence the divergent curvature of the phase boundary $\rho$ near the CEP can be expressed in terms of the amplitudes $X_\pm$, $Y$, $Z_\pm$. The ratio of these amplitudes is related to the universal critical amplitude ratio of the critical phase boundary $\lambda$ (see Sec. \ref{sec2} Eq. \ref{eqq1})
\begin{eqnarray}
 \frac{X_ +}{X_ -} = \frac{A_ +}{A_ -} \\
 \frac{Z_ +}{Z_ -} = \frac{C_ +}{C_ -} \\
  \frac{X_+ Z_+}{Y^2}= \frac{A_+ C_+}{(1- \alpha)(2- \alpha) B^2}
\end{eqnarray}

 Since the critical amplitude ratios $\frac{A_ +}{A_ -}$ and $\frac{C_ +}{C_ -}$ are universal, the ratios $\frac{X_ +}{X_ -}$ and $\frac{Z_ +}{Z_ -}$  are also universal. Therefore, the singularity in $\Delta_{\rho} (T, H)$  of the phase boundary $\rho$ can be verified by calculating the derivatives   $\frac{\partial \, \Delta_{\rho} (T, H)}{\partial T}$, $\frac{\partial^2 \, \Delta_{\rho} (T, H)}{\partial T^2}$, $\frac{\partial \, \Delta_{\rho} (T, H)}{\partial H}$ and  $\frac{\partial^2 \, \Delta_{\rho} (T, H)}{\partial H^2}$ near the CEP.

\subsection{Verification of the singularities of the phase boundary near a CEP}
In this section, we verify the Fisher \textit{et al} scaling argument near the CEP by studying the curvature of the phase boundary $\Delta_{\rho} (T, 0)$ i.e. $\frac{\partial ^2 \Delta}{\partial T^2}$, for RCFBC model (for the range $ 0.022 < p \leq 0.107875$) and for the RBEG model (for the range ``$ - 0. 1838 < K \leq - 1$"). To contrast, we also show their behaviour  near a TCP.

\begin{figure}
\centering
     \begin{subfigure}[b]{0.49\textwidth}
         \centering
         \includegraphics[width=\textwidth]{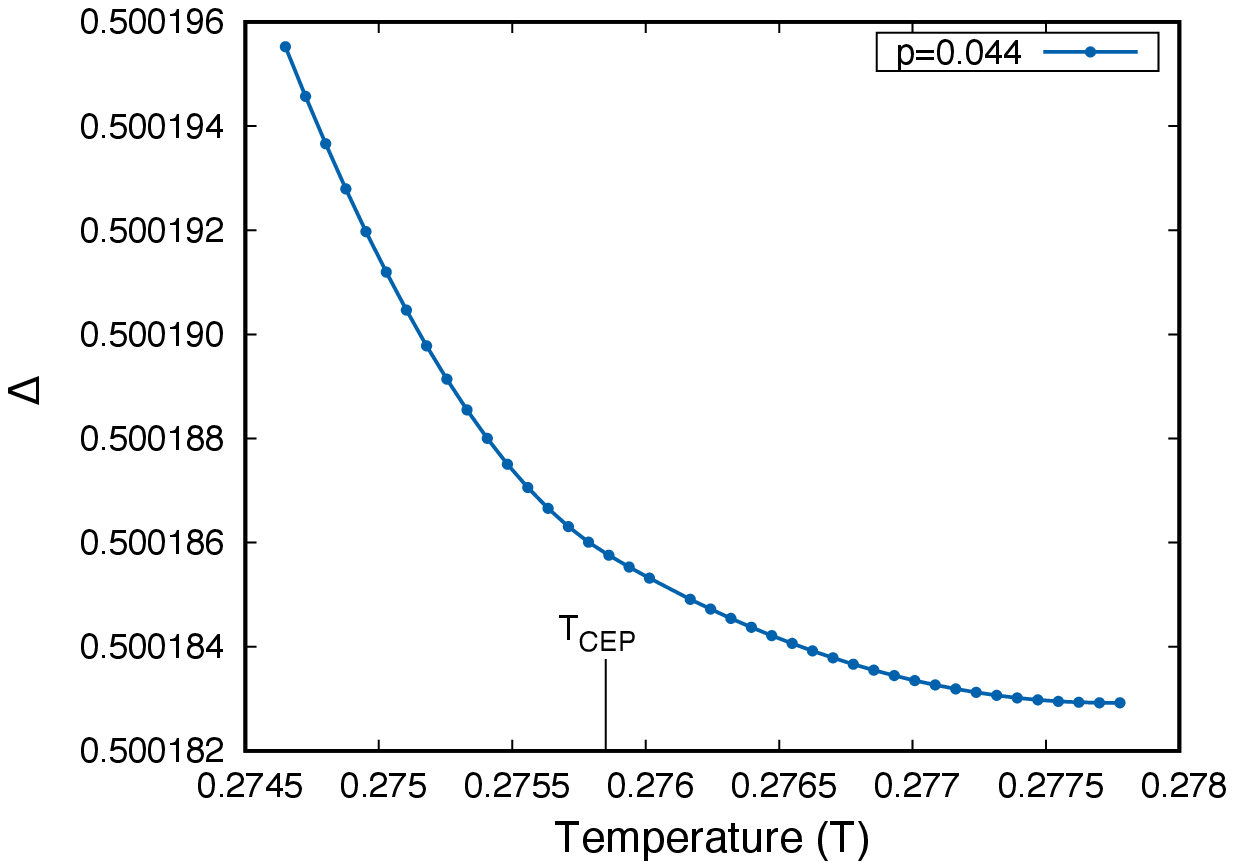}
         \caption{}
         \label{fig560}
     \end{subfigure}
     \hfill
     \centering
     \begin{subfigure}[b]{0.49\textwidth}
         \centering
         \includegraphics[width=\textwidth]{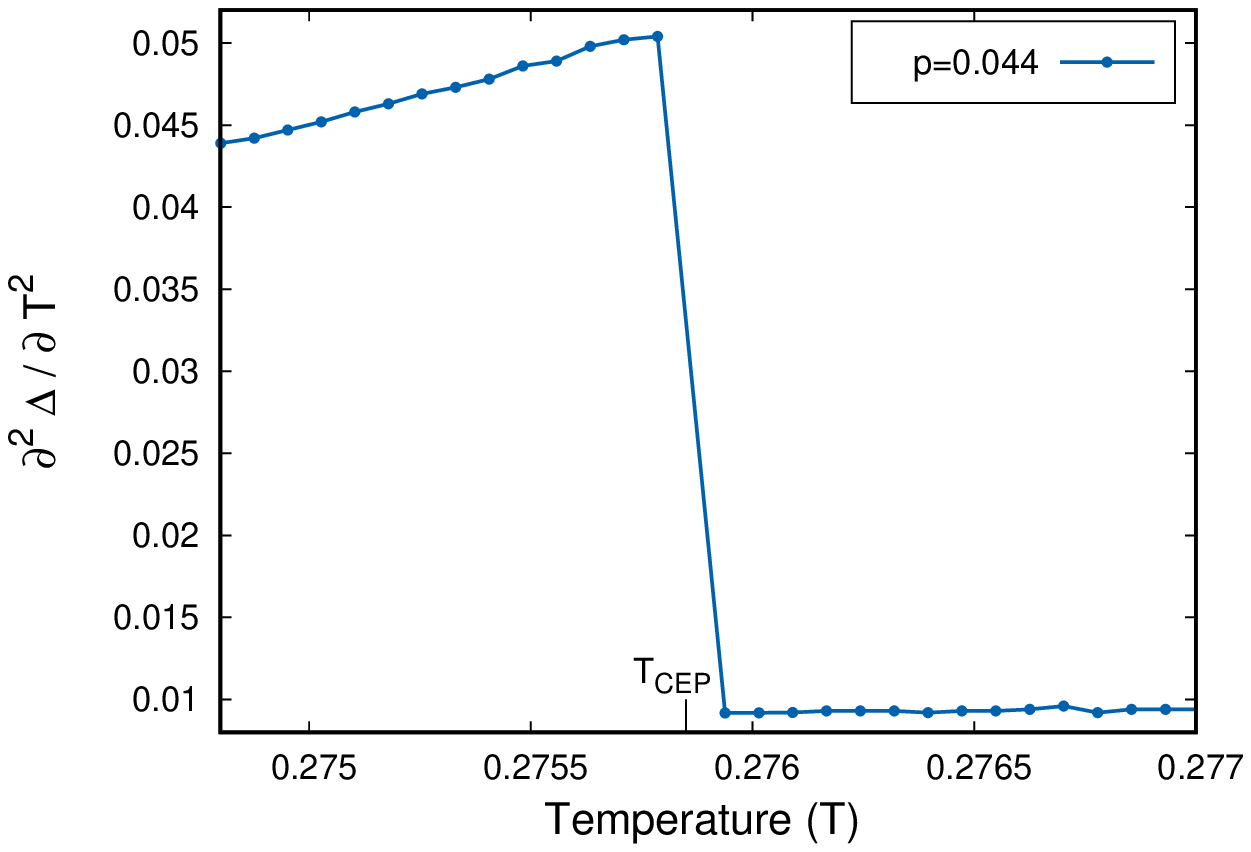}
         \caption{}
         \label{fig561}
     \end{subfigure}
     \hfill
     \begin{subfigure}[b]{0.49\textwidth}
         \centering
         \includegraphics[width=\textwidth]{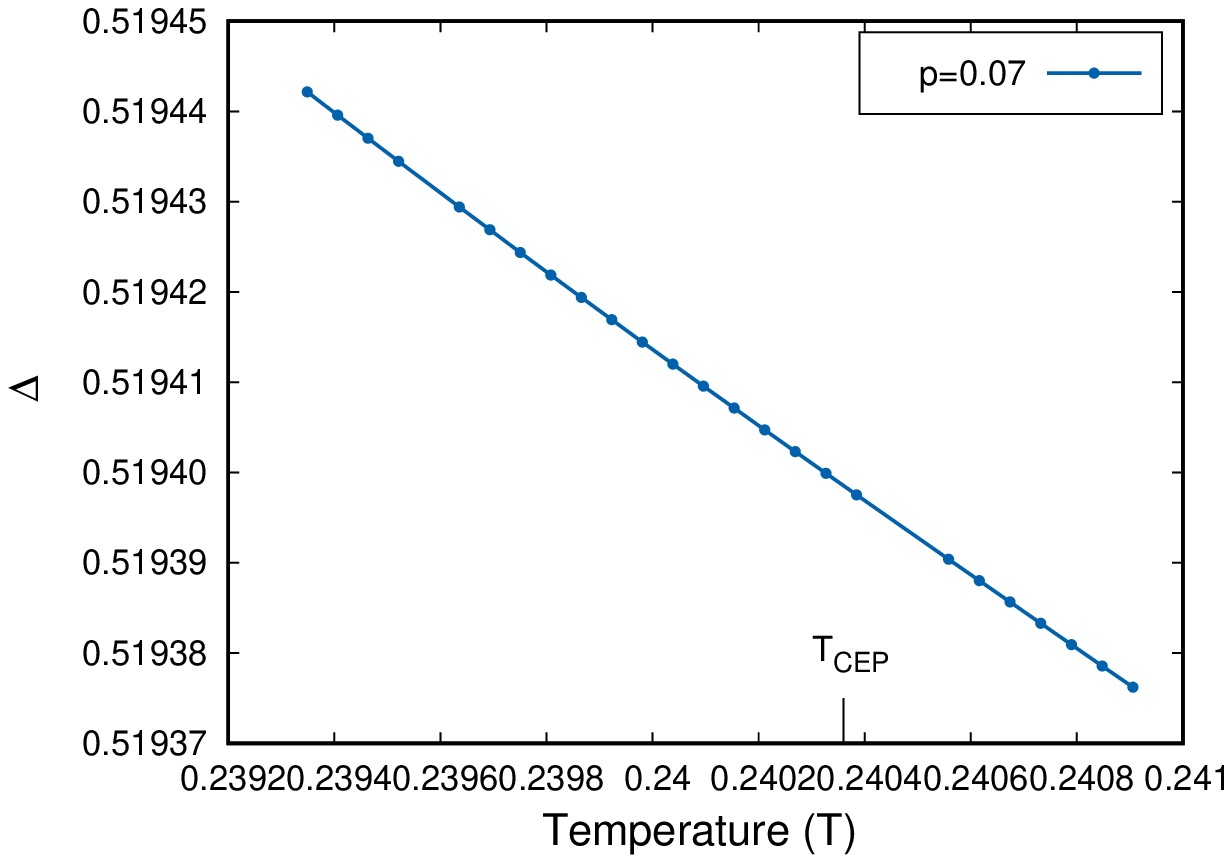}
         \caption{}
         \label{fig562}
     \end{subfigure}
     \hfill
     \centering
     \begin{subfigure}[b]{0.49\textwidth}
         \centering
         \includegraphics[width=\textwidth]{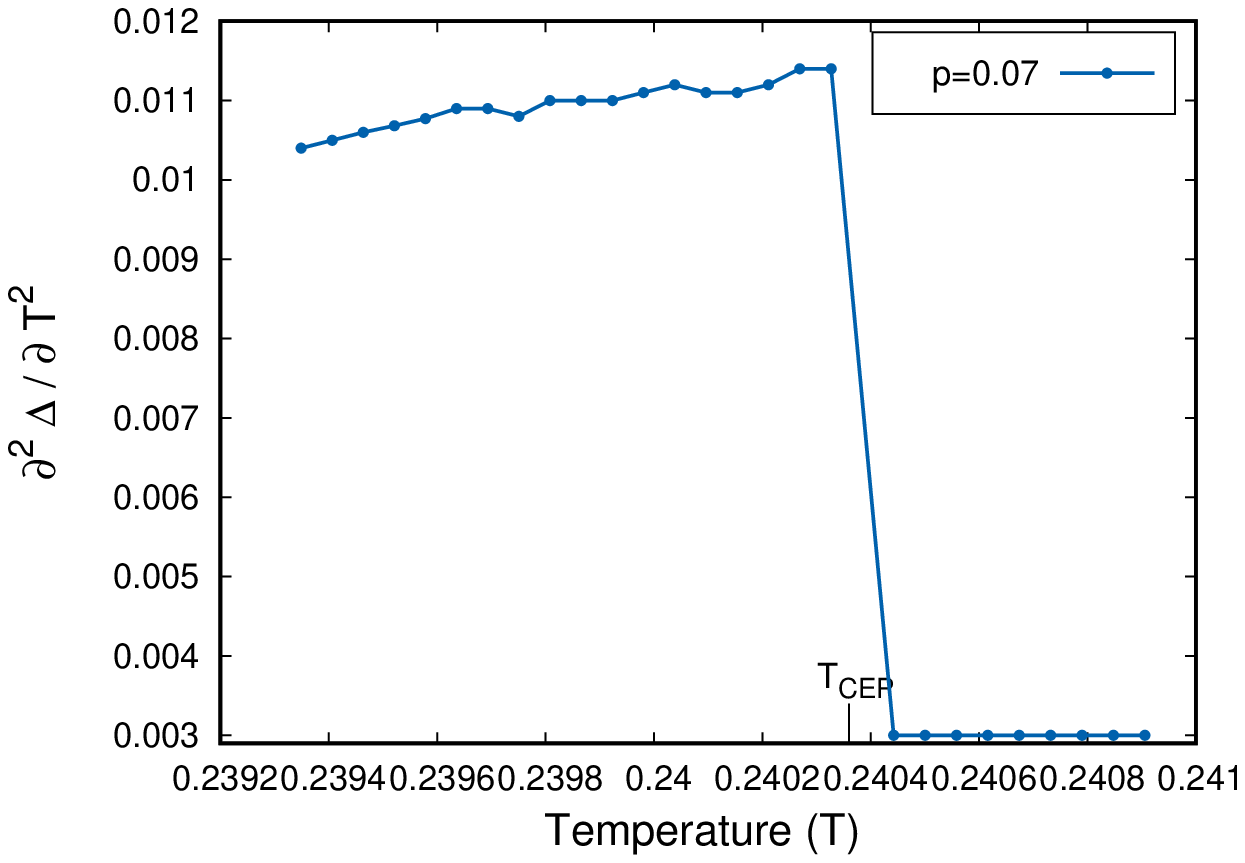}
         \caption{}
         \label{fig563}
     \end{subfigure}
     \hfill
     \begin{subfigure}[b]{0.49\textwidth}
         \centering
         \includegraphics[width=\textwidth]{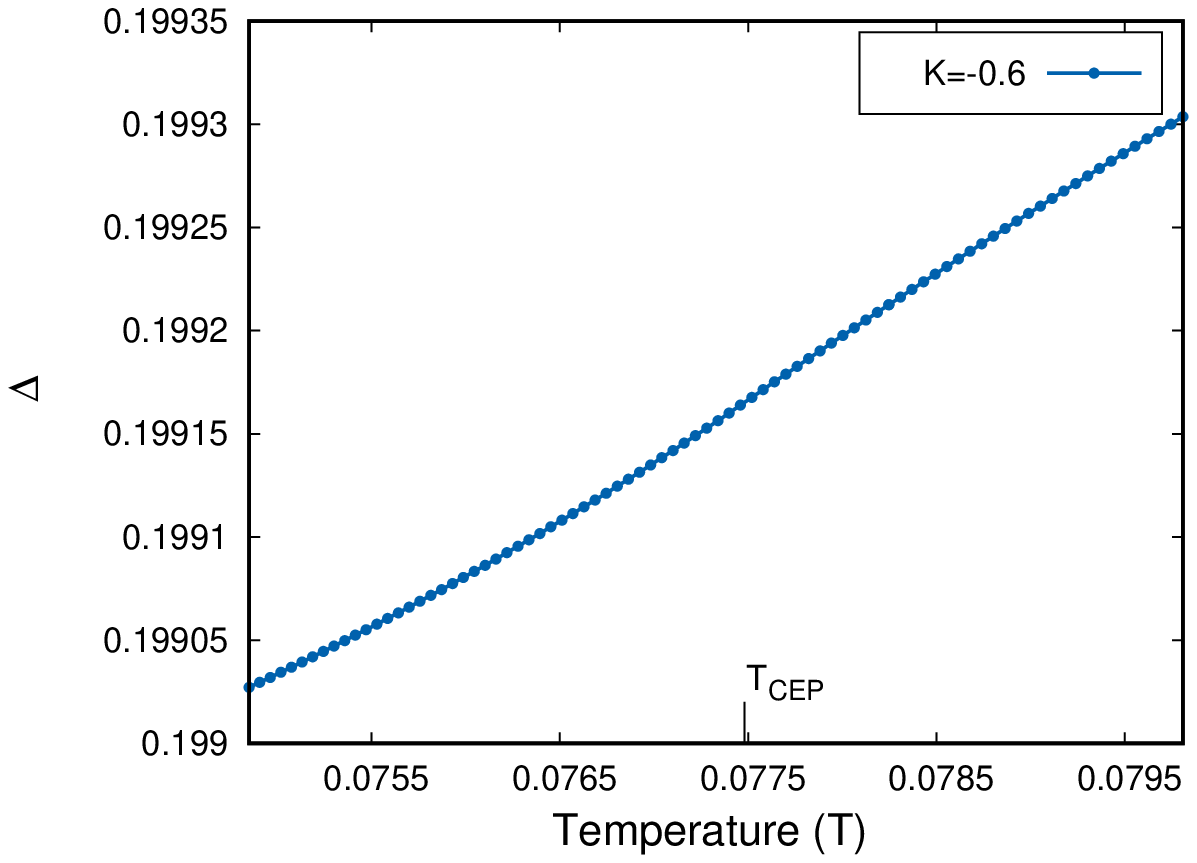}
         \caption{}
         \label{fig600}
     \end{subfigure}
     \hfill
     \centering
     \begin{subfigure}[b]{0.49\textwidth}
         \centering
         \includegraphics[width=\textwidth]{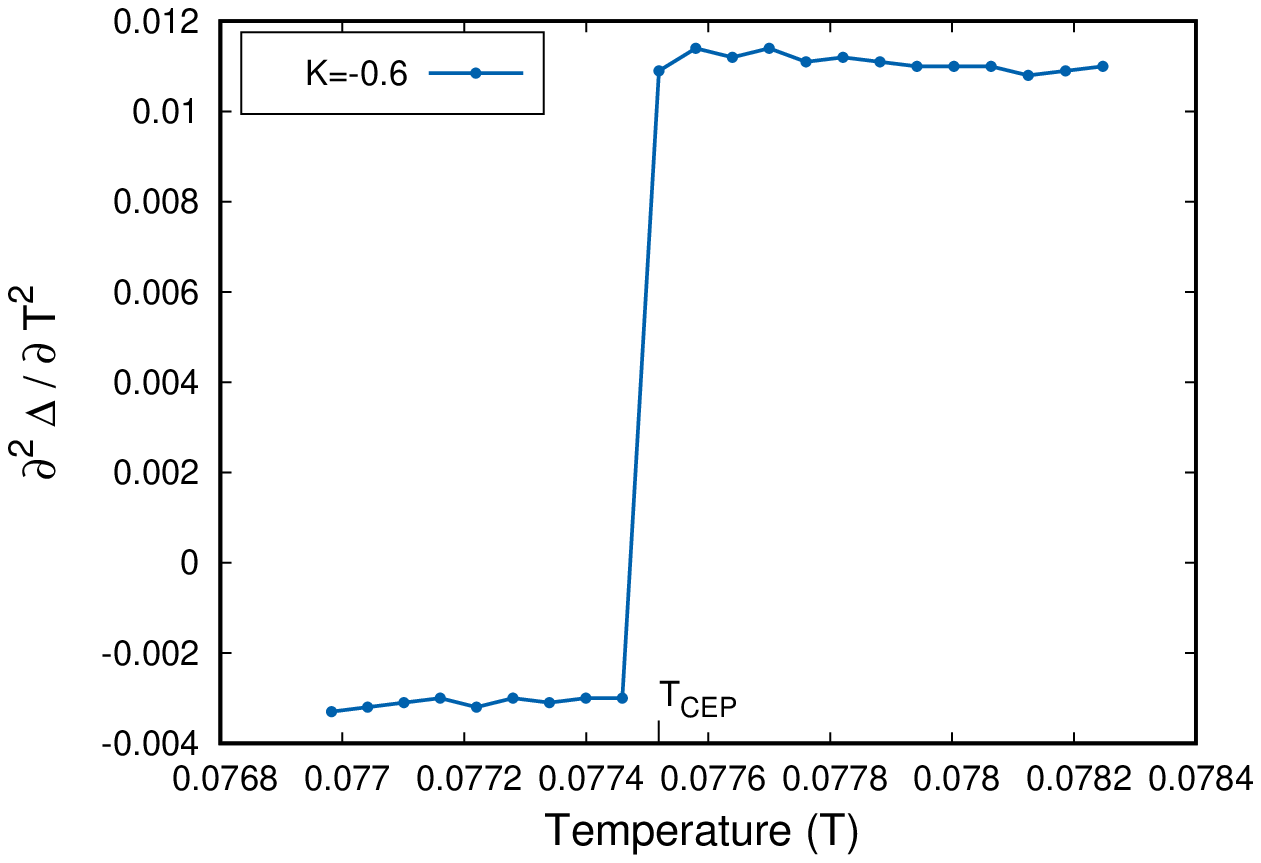}
         \caption{}
         \label{fig601}
     \end{subfigure}
     \hfill
        \caption{The phase boundary curvature as a function of temperature near a CEP for RCFBC model at $p=0.044$ and $0.07$, and for RBEG model  at $K=-0.6$. Fig. \textbf{(a)},  \textbf{(c)} and \textbf{(e)} show the plot of the spectator phase boundary $\Delta_{\rho} (T, 0)$  as a function of $T$. Fig.  \textbf{(b)},  \textbf{(d)} and \textbf{(f)} show the plot of the second derivative of the phase boundary $\frac{\partial ^2 \Delta}{\partial T^2}$ as a function of $T$. }
        \label{fig56}
\end{figure}

 In order to observe the singularity in the phase boundary $\rho$, we obtain the co-ordinates  $\Delta_{\rho} (T, 0)$ of the line of co-existence  for different values of $T$ by equating  $f(m)$ for both phases keeping $H=0$. We then obtain its second order derivative $\frac{\partial ^2 \Delta_{\rho}(T, 0)}{\partial T^2}$ using numerical differentiation. The $\Delta_{\rho}(T, 0)$,  $\frac{\partial ^2 \Delta_{\rho}(T, 0)}{\partial T^2}$ are plotted in Fig. \ref{fig56} for RCFBC model for  $p=0.044$, $p=0.07$  and for RBEG model for $K=-0.6$.

\begin{figure}
    \centering
 \begin{subfigure}[b]{0.49\textwidth}
         \centering
         \includegraphics[width=\textwidth]{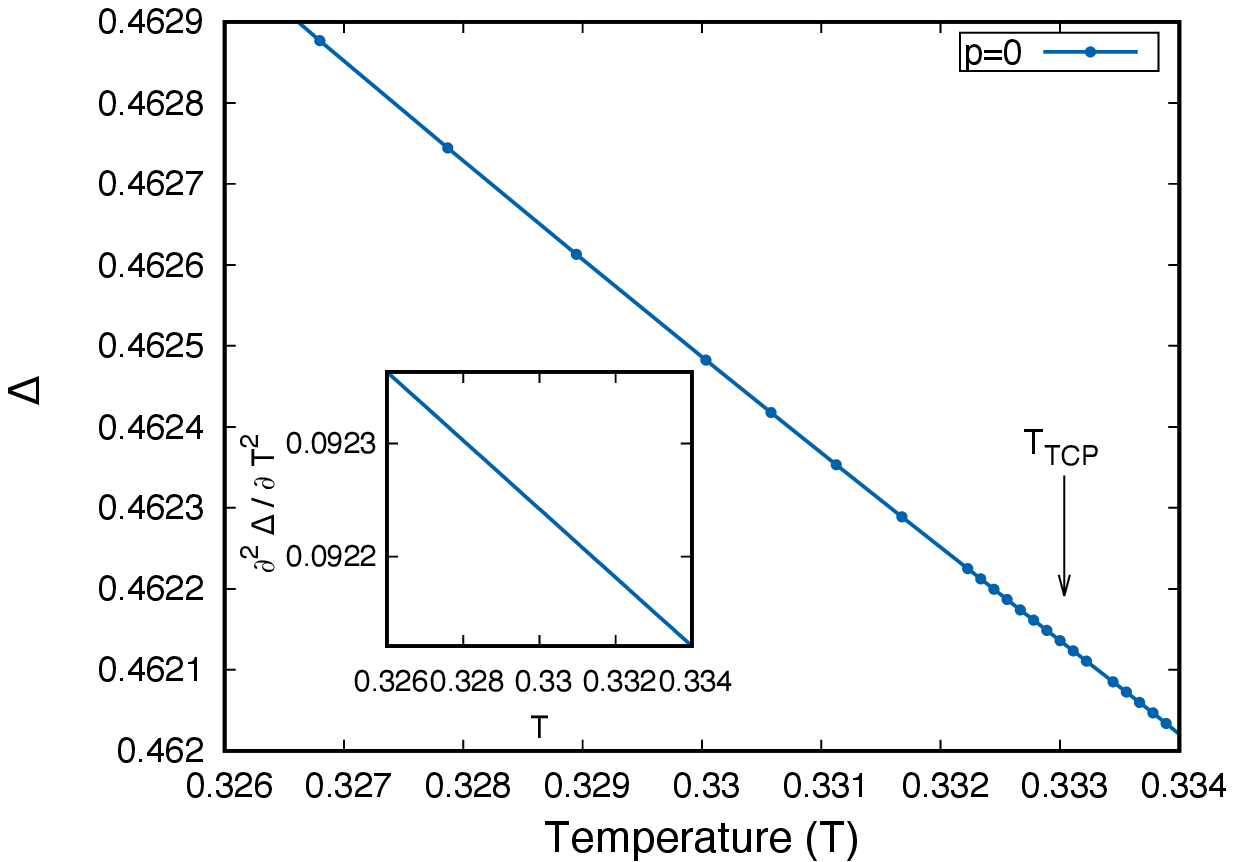}
         \caption{}
         \label{fig590}
     \end{subfigure}
     \hfill
     \centering
     \begin{subfigure}[b]{0.49\textwidth}
         \centering
         \includegraphics[width=\textwidth]{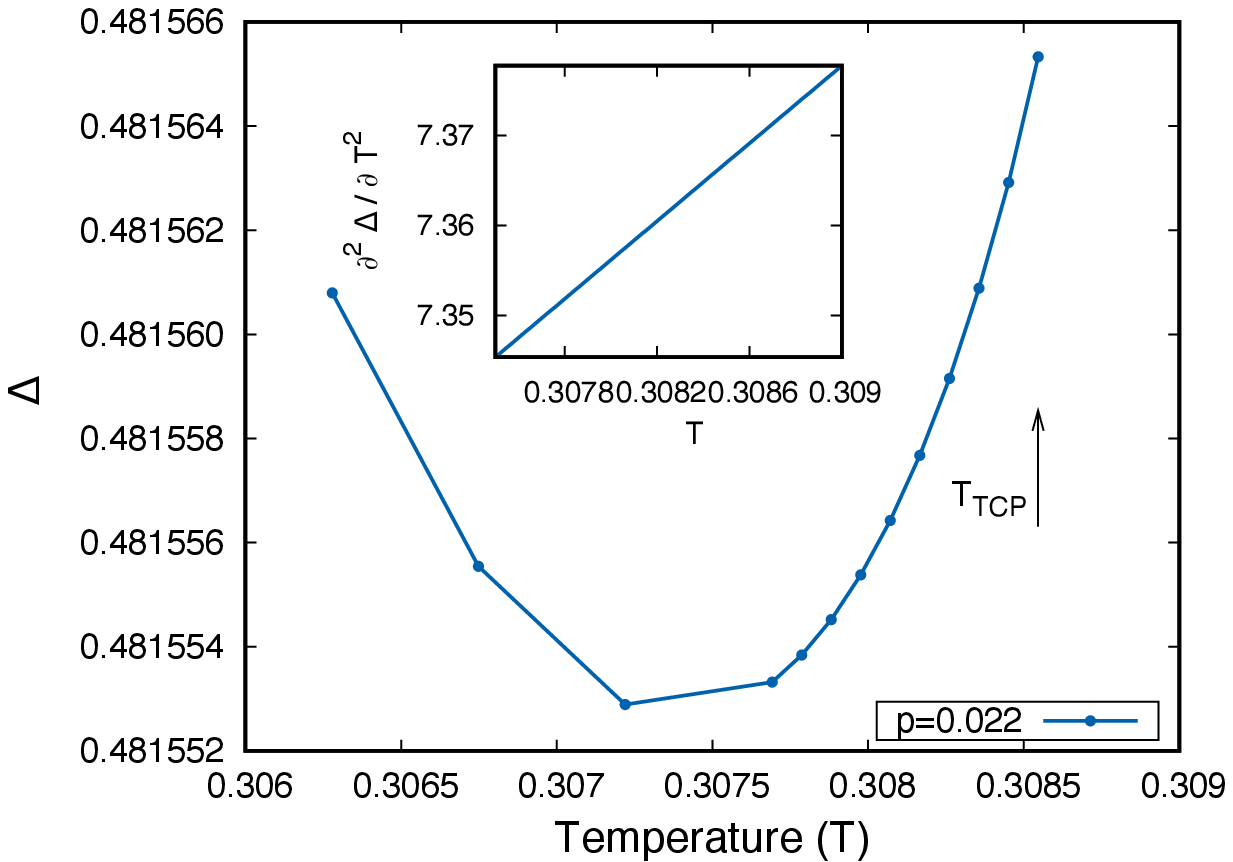}
         \caption{}
         \label{fig591}
     \end{subfigure}
     \hfill
        \caption{The phase boundary $\Delta_{\rho} (T, 0)$ as a function of  $T$ in order to compare the scaling arguments near a TCP for RCFBC model at $p=0$ and $p=0.022$. We plot the co-ordinate of $\Delta_{\rho}(T, 0)$ for \textbf{(a)} $p=0$ and \textbf{(b)} $p=0.022$. Near the TCP, the $\Delta_{\rho} (T, 0)$ changes smoothly with $T$ and  the  second derivative, $\frac{\partial ^2 \Delta}{\partial T^2}$ does not exhibit any singularity  for both the values of $p$, shown in the insets. }
        \label{fig58new}
\end{figure}

 In Fig. \ref{fig560}, \ref{fig562}, \ref{fig600} we plot the co-ordinates of the phase boundary $\Delta$ as a function of $T$.  In Fig. \ref{fig561}, \ref{fig563}, \ref{fig601}  we plot the second derivative of the phase boundary $\frac{\partial ^2 \Delta }{\partial T^2}$ as a function of $T$. We find that the phase boundary shows a discontinuity at the $T_{CEP}$. The discontinuity observed in the $\frac{\partial ^2 \Delta }{\partial T^2}$ is similar to the jump in the specific heat plot ($C_v$) as a function of $T$ along the $\lambda$ line. This is expected from Eq. \ref{eq-fisher} as the curvature of the phase boundary  $\Delta_{\rho} (T, 0)$ near the CEP in the absence of the external field $H$   diverges as
 
 \begin{equation}
     \frac{\partial ^2 \Delta_{\rho} (T, 0) }{\partial T^2} =  X_\pm  |\hat{t}|^{- \alpha} \,\,\,\,\,\,\,\,   \text{with $\alpha = 0$}
 \end{equation}
and the ratio of the amplitudes  $X_+/ X_-$ is given by the ratios of the values of $\frac{\partial ^2 \Delta_{\rho} (T, 0) }{\partial T^2}$ at $T_{CEP}$.

We  also plot the  phase boundary $\Delta_{\rho}(T, 0)$ as a function of $T$ near a TCP. In Fig. \ref{fig590} and Fig. \ref{fig591} we plot them for $p=0$ and $p=0.022$ respectively for the RCFBC model. In Fig. \ref{fig590}, the $\Delta$ along the co-existence region changes continuously for $p=0$. For $p=0.022$, $\Delta$ shows a non-monotonicity. But in both cases, the plot is smooth near a TCP. And the second derivative of the $\Delta$ as a function of $T$ does not show any singularity  near a TCP (see the insets in both the plots where $\frac{\partial ^2 \Delta_{\rho} (T, 0) }{\partial T^2}$ changes continuously with $T$). This behaviour  is in contrast with Fig. \ref{fig561}, \ref{fig563} for the similar quantities near a CEP, where the second derivative shows a discontinuity.

\begin{figure}
     \centering
     \begin{subfigure}[b]{0.49\textwidth}
         \centering
         \includegraphics[width= \textwidth]{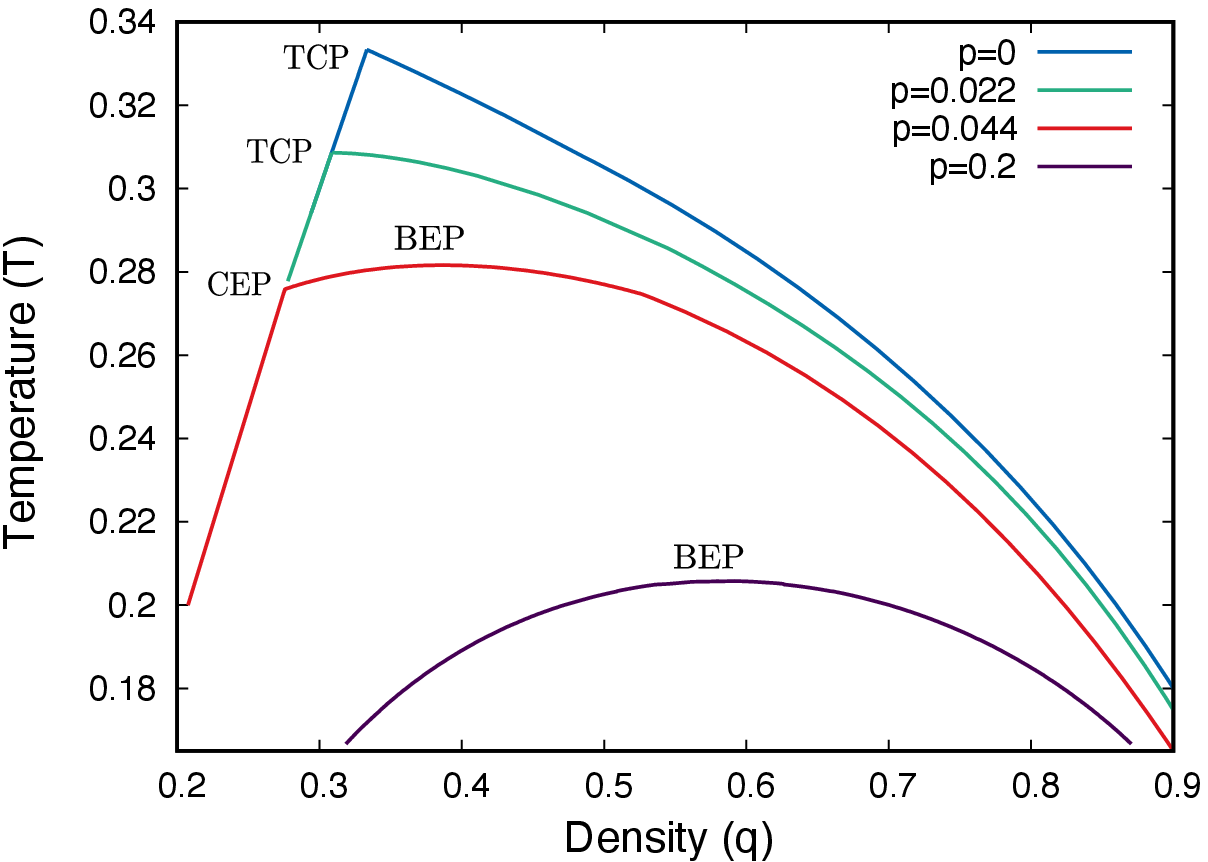}
         \caption{ }
        \label{fig620}
     \end{subfigure}
     \hfill
     \begin{subfigure}[b]{0.49\textwidth}
         \centering
        \includegraphics[width= \textwidth]{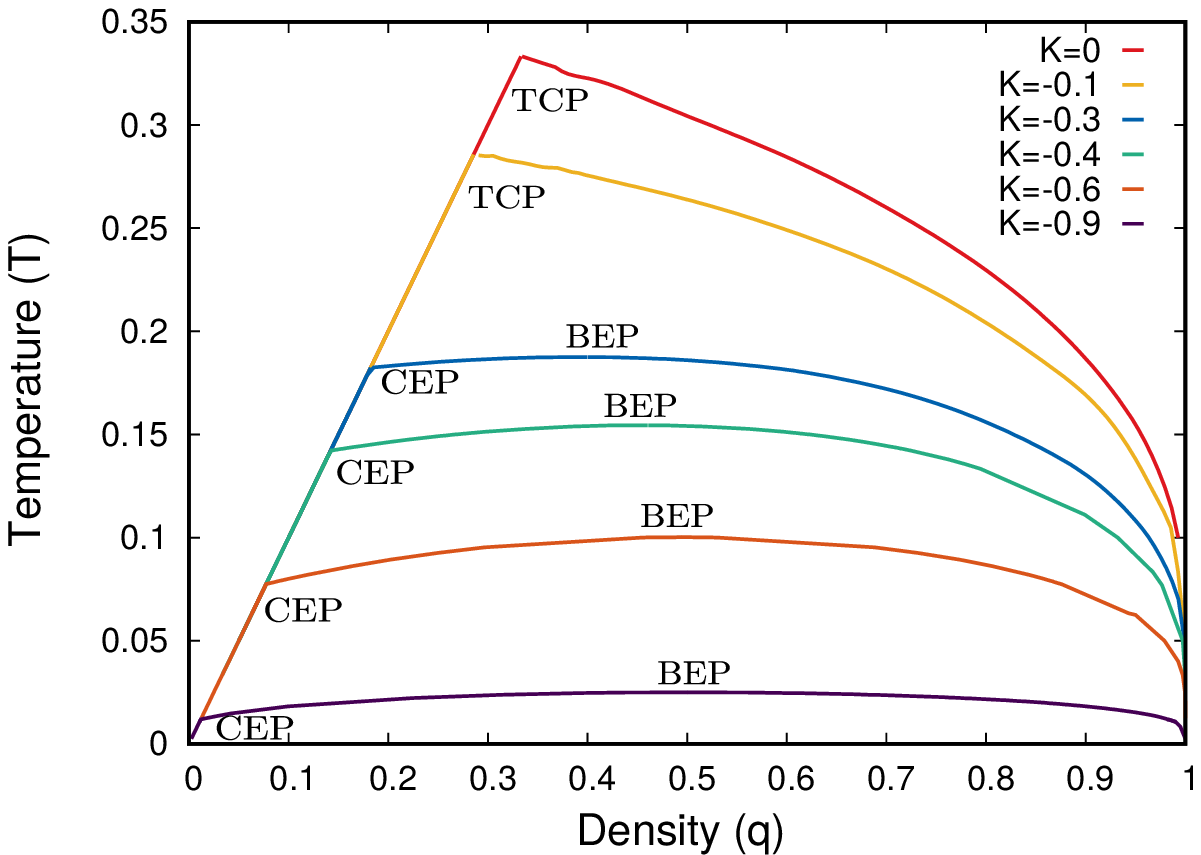}
         \caption{}
         \label{fig61}
    \end{subfigure}
   \hfill
    \caption{ The behaviour of the phase co-existence region near a CEP and a TCP. Plot of the  $T$ - $q$ plane  \textbf{(a)} for RCFBC model for a range of $p$ and  \textbf{(b)} for  RBEG model for a range of $K$.}
       \label{fig69}
\end{figure}


\section{Study of the co-existence curves} \label{sec5}

Both the TCP and CEP are points where  $\lambda$ line meets the  phase  co-existence envelope.  In order to distinguish between these two multicritical points we find that it is illustrative to look at the co-existence plot in the $T-q$ plane. We find that the shape of the phase co-existence curve is different for a CEP and a TCP. In Fig. \ref{fig69} we plot the phase co-existence curve for the RCFBC model and RBEG model for a range of $p$ and $K$ respectively by comparing the free energies of the phases for different values of $T$ and $\Delta$. Depending on the values of $p$ and $K$, the $\lambda$ line meets the boundary of the co-existence curve at a TCP or at a CEP (the $\lambda$ line in the $T-q$ plane is just a straight line $q=T$, given in Eq. \ref{eqq1}). 

We observe that  for all the ranges of the parameters ($0 \leq p \leq 0.022$ for RCFBC model shown in Fig. \ref{fig620} and $-0.1838 \leq  K \leq 0$ for RBEG model shown in Fig. \ref{fig61}) for which the TCP exists, the $\lambda$ line meets the  peak of the co-existence region, where the phase co-existence region goes to zero. As the parameters change ($0.022 < p \leq 0.107$ for RCFBC model and $-1 \leq  K < -0.1838$ for RBEG model),   the $\lambda$ line does not terminate at the peak of the co-existence region. The peak is now a BEP and the $\lambda$ line truncates on either side of the co-existence region, giving rise to a kink at the CEP. In RCFBC model for $0.107 < p \leq 0.5$, the co-existence region moves away from the $\lambda$ line, and the $\lambda$ line terminates at $T=p$ (shown in Fig. \ref{fig620} for $p=0.2$). For $K < -1$ in RBEG model, the BEP and CEP, along with the co-existence region move to $T=0$ and only the $\lambda$ line remains. 

\begin{figure}
     \centering
         \includegraphics[scale=0.8]{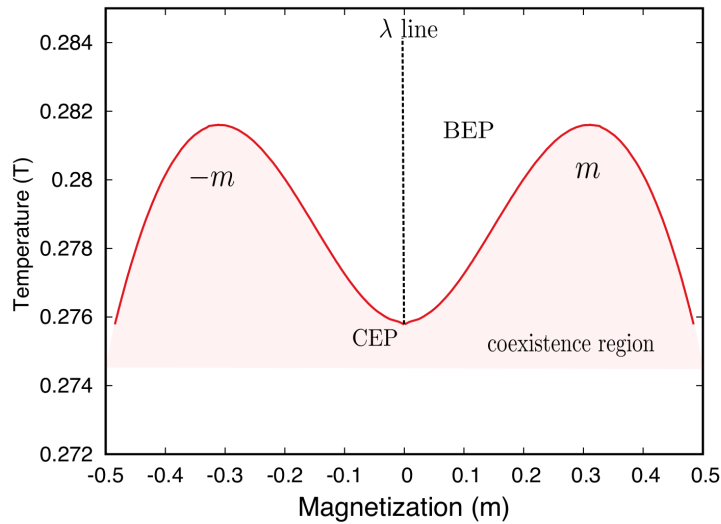}
    \caption{ Temperature ($T$) vs. magnetization ($m$) plot near a BEP and CEP for RCFBC model at $p=0.044$. The area under the red line is the magnetization co-existence region.  The BEP is a point of co-existence of two critical phases ($\pm m$) in the $T-m$ plane. CEP is the point where the $\lambda$ line terminates at the co-existence region. The co-existence curve near the CEP doesn’t show any singularity in the $T-m$ plane. }
      \label{fig76new}
 \end{figure}

 In the $T-q$ co-existence  plane, the BEP appears like a CP where the density co-existence region ends. But the fact that the BEP is different than a CP  can be easily seen from the  $T$-$m$ plot. Fig. \ref{fig76new} shows that the BEP is a  point of co-existence of two critical phases ($\pm m$). The singularity seen in the  $T-q$ co-existence plane near a CEP vanishes in the $T-m$ co-existence plane.

 \begin{figure}
\centering
     \begin{subfigure}[b]{0.49\textwidth}
         \centering
         \includegraphics[width=\textwidth]{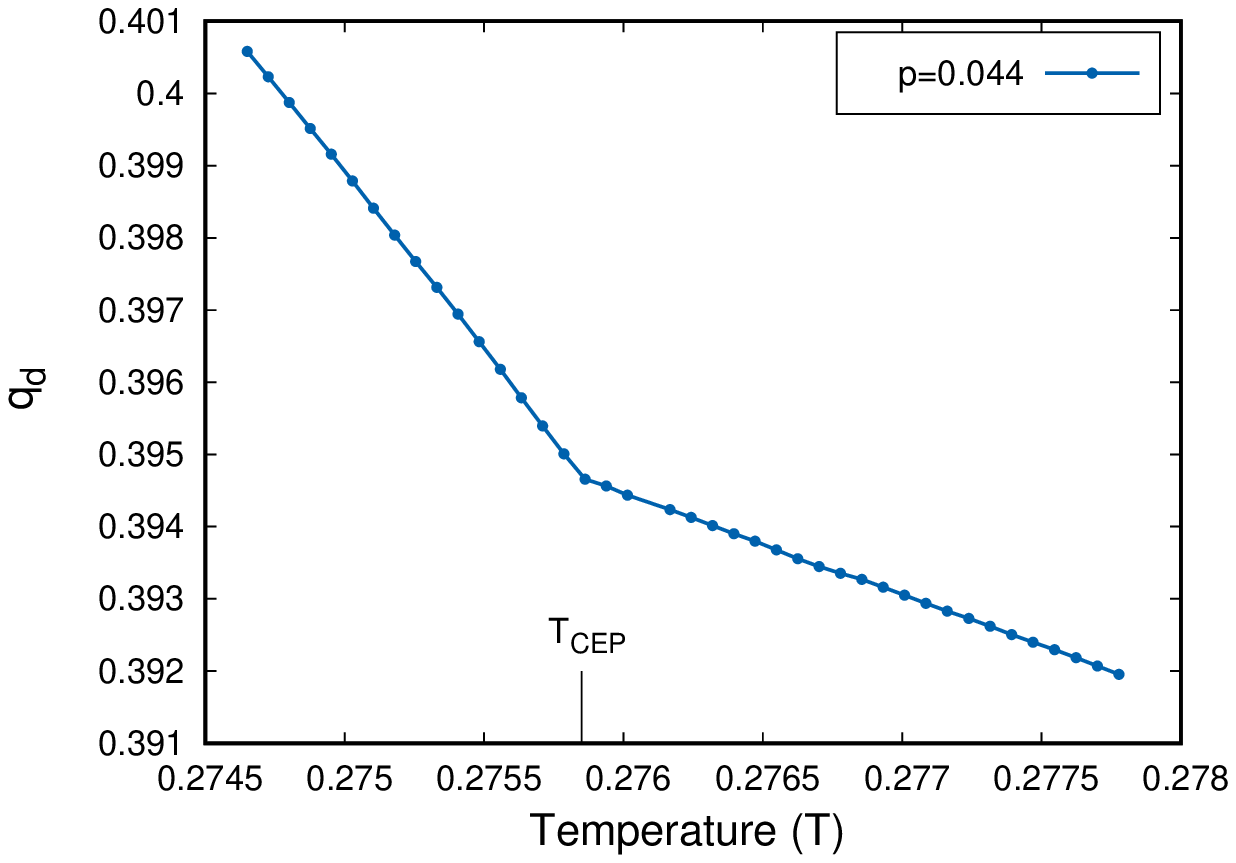}
         \caption{}
         \label{fig570}
     \end{subfigure}
     \hfill
     \centering
     \begin{subfigure}[b]{0.49\textwidth}
         \centering
         \includegraphics[width=\textwidth]{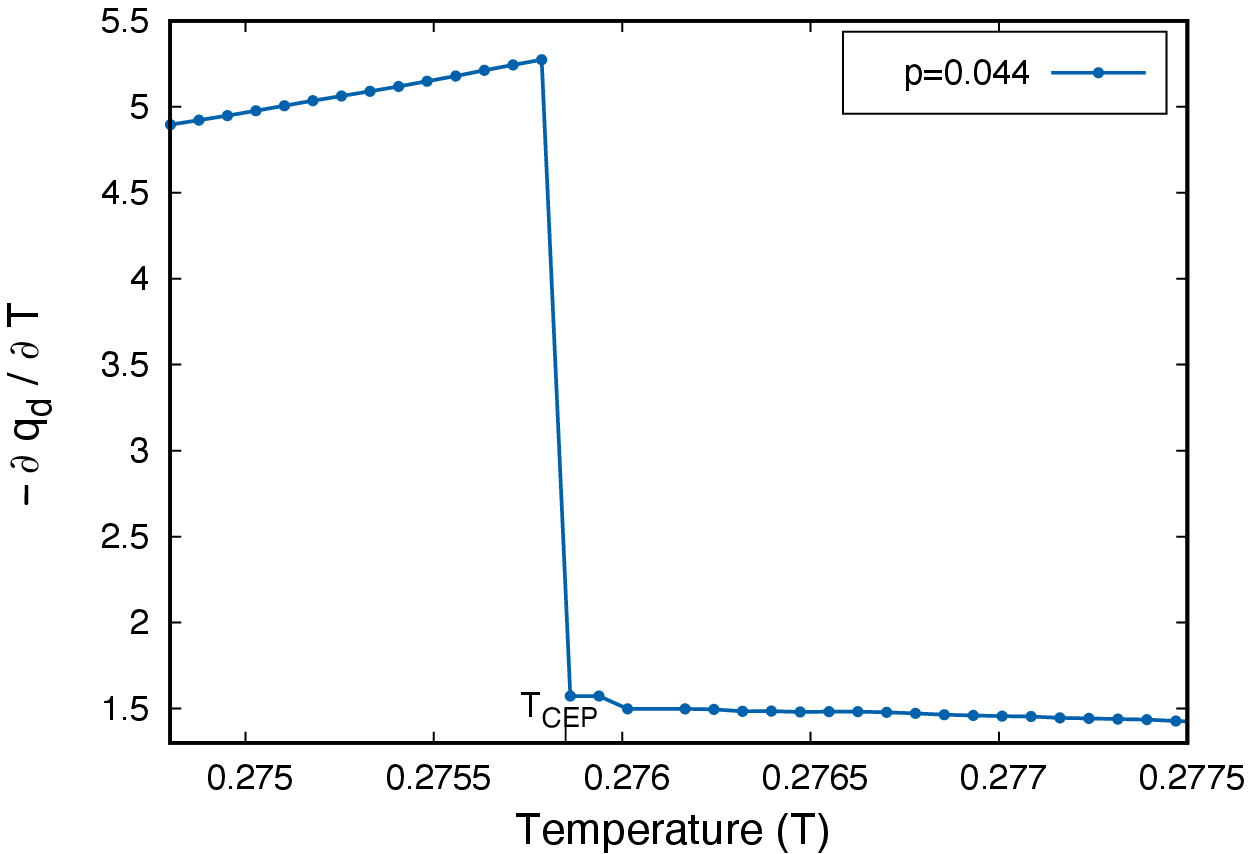}
         \caption{}
         \label{fig571}
     \end{subfigure}
     \hfill
     \begin{subfigure}[b]{0.49\textwidth}
         \centering
         \includegraphics[width=\textwidth]{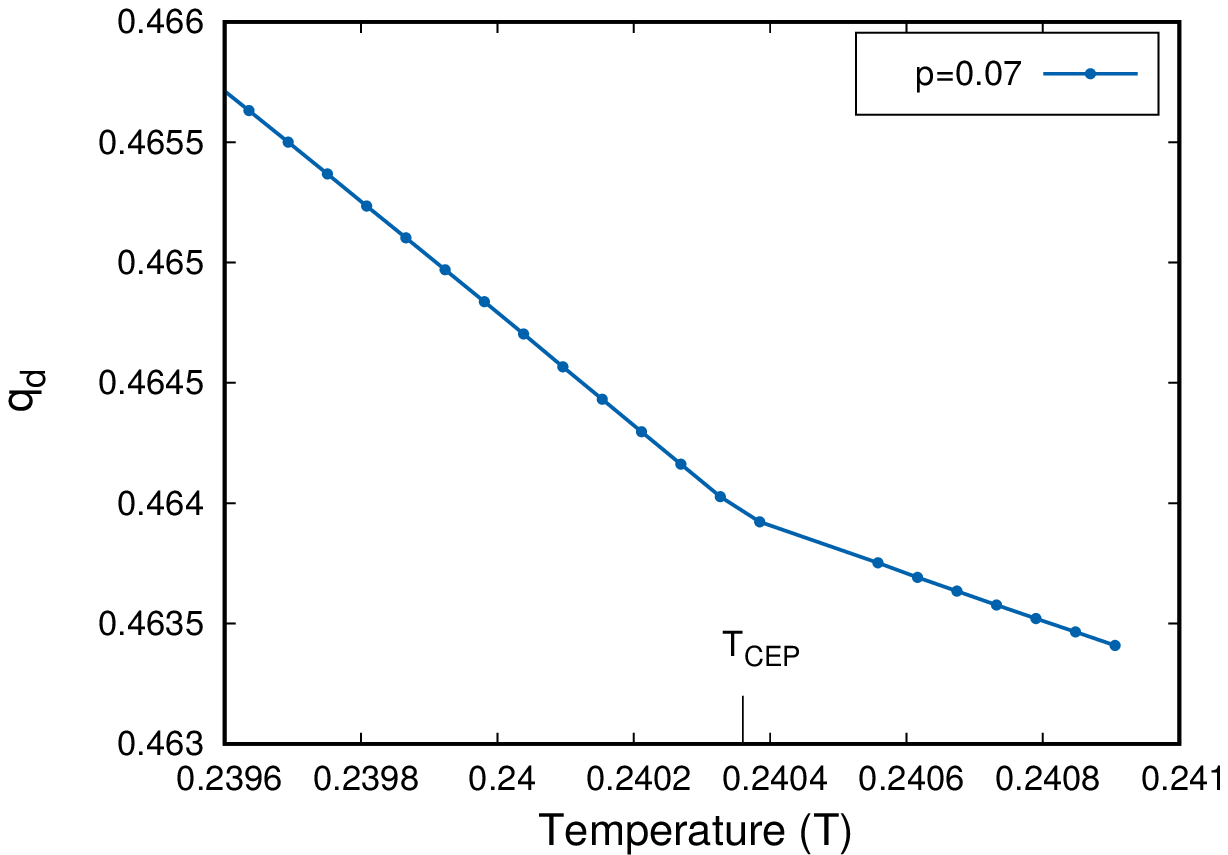}
         \caption{}
         \label{fig572}
     \end{subfigure}
     \hfill
     \centering
     \begin{subfigure}[b]{0.49\textwidth}
         \centering
         \includegraphics[width=\textwidth]{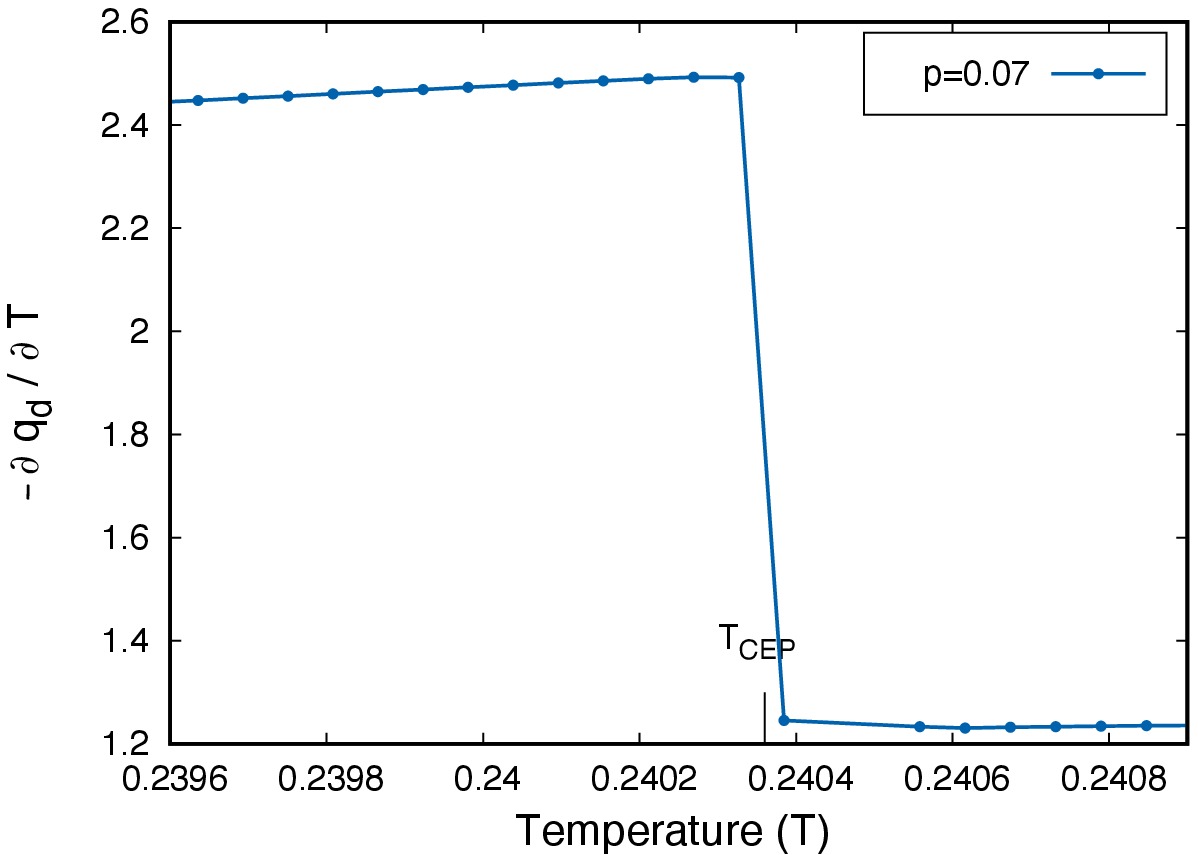}
         \caption{}
         \label{fig573}
     \end{subfigure}
     \hfill
     \begin{subfigure}[b]{0.49\textwidth}
         \centering
         \includegraphics[width=\textwidth]{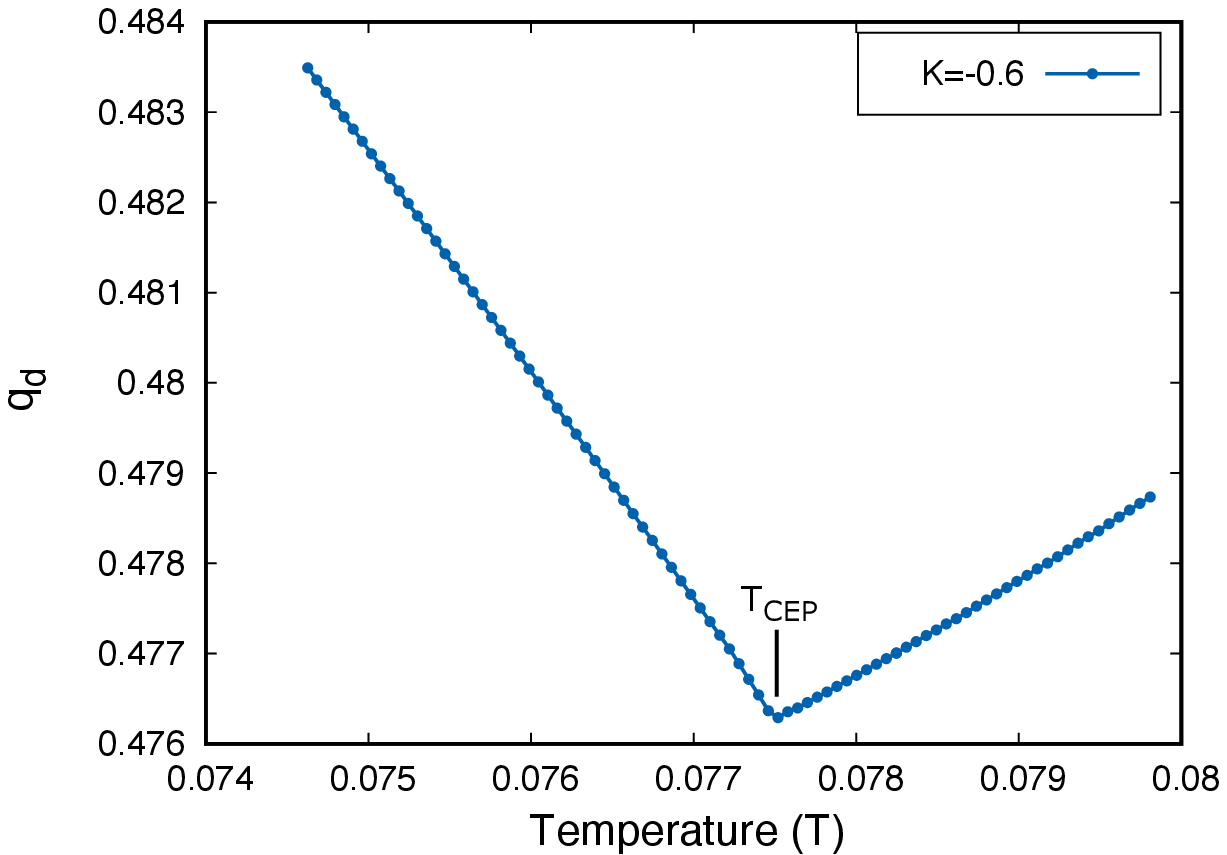}
         \caption{}
         \label{fig602}
     \end{subfigure}
     \hfill
     \centering
     \begin{subfigure}[b]{0.49\textwidth}
         \centering
         \includegraphics[width=\textwidth]{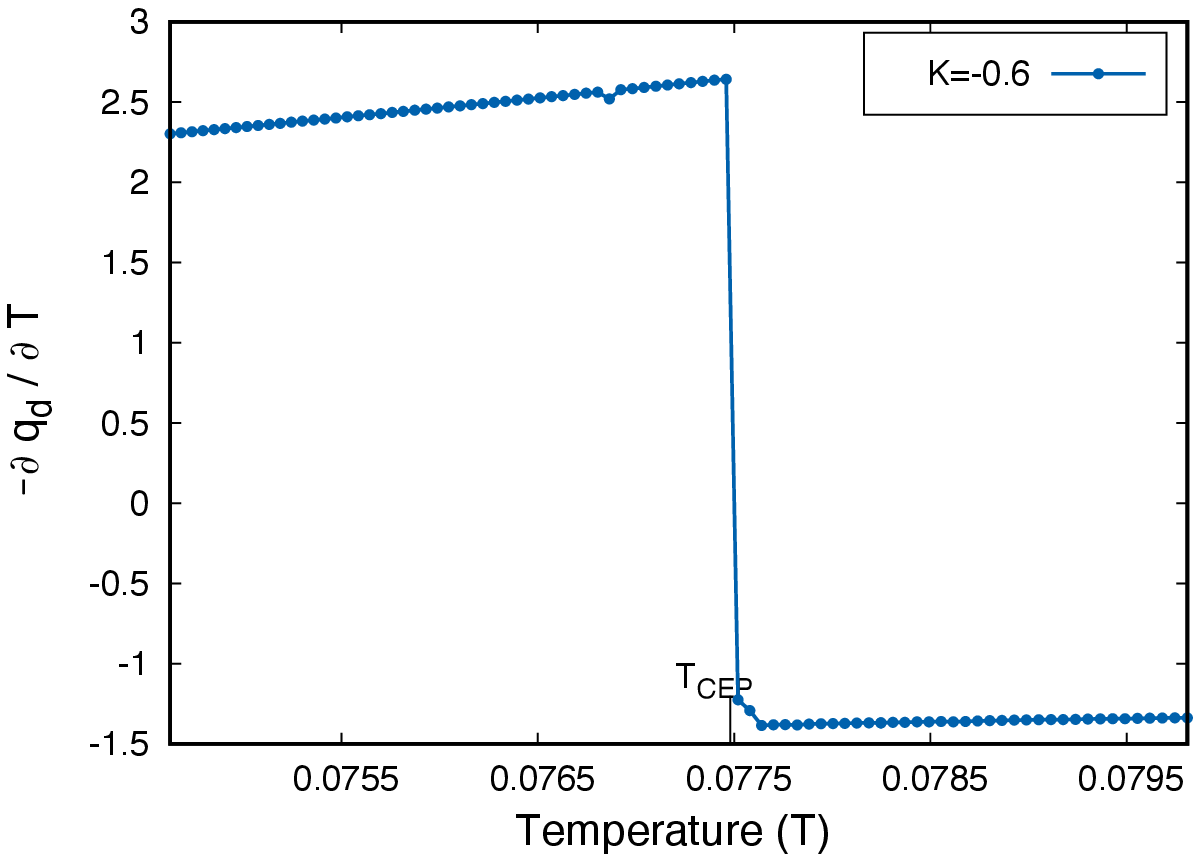}
         \caption{}
         \label{fig603}
     \end{subfigure}
     \hfill
        \caption{The discontinuity of the phase co-existence curve near a CEP for RCFBC model at $p=0.044$ and $p=0.07$, and for RBEG model at $K=-0.6$. Fig. \textbf{(a)}, \textbf{(c)} and \textbf{(e)} show the plot of the co-existence diameter ($q_d$)  as a function of  $T$. Fig.  \textbf{(b)}, \textbf{(d)} and \textbf{(f)} show the plot of the negative derivative of the diameter $-\frac{\partial q_{q}}{\partial T}$ as a function of $T$.  }
        \label{fig57}
\end{figure}


\subsection{Singularities in the phase co-existence curves }

B. Wilding extended the Fisher's scaling argument \cite{PhysRevLett.78.1488, PhysRevE.55.6624} in order to study the singularities in the thermodynamic density $q$  conjugate to $\Delta$. It was shown for a binary fluid mixture that in addition to the singularity of the phase boundary $\Delta_\rho(T,H)$ proposed by Fisher \textit{et al}, there are other singularities in the diameter of the  co-existence curve. The diameter is the  order parameter ($q$) conjugate to the non-ordering field ($\Delta$). The density $q$ can be obtained from the free energy $f ( T, H, \Delta)$ as
\begin{equation}\label{eqq}
    q = - \frac{1}{V} \,\, \Bigg ( \frac{\partial f ( T, H, \Delta)}{\partial \Delta} \Bigg )_{T, H}
\end{equation}

Wilding showed that the temperature derivative of the co-existence diameter  diverges at the CEP. The diameter of the co-existence curve is defined as

\begin{equation}\label{eqqd}
 q_d (T) \equiv \frac{1}{2} \Big (q_{p/m_2} (\Delta_{\rho} (T)) + q_{m_1} (\Delta_{\rho} (T)) \Big )
\end{equation}
here $q_{m_1}$, $q_{p/m_2}$ are the densities  of the  phases $m_1$ and the phases $p/m_2$  respectively (see Fig. \ref{fig610}).  Using the Eqs.  \ref{eqq}, \ref{eqqd} and the phenomenological  scaling theory, the singular behaviour of the diameter $q_d$ can be written as
\begin{equation}
    q_d (T) \approx  U_\pm  |\hat t|^{1 - \alpha} + \text{terms analytic at $T_{CEP}$}
\end{equation}
 where the non-universal amplitudes $U_\pm$ can be expressed in terms of the scaling function $W_\pm$ (Eq. \ref{criticaleq}) as follows
 \begin{equation}
     U_\pm = b_2 \,\,  (2 - \alpha) (1+ b_2 \Delta_1)^{1- \alpha} \,\,\,  W_\pm(0)
 \end{equation}
here $b_2$ is the gradient  of the $\lambda$ line calculated at the CEP : 
$ b_2 = - \pdv{T_c(\Delta)}{\Delta}$ and $\Delta_1$ is the non-singular coefficient given in Eq. \ref{eq-fisher}.

Thus the scaling argument shows that the co-existence curve diameter $q_d(T)$ exhibits a divergence near the CEP and it can be verified by plotting the derivative of the $q_d(T)$
\begin{equation}
\pdv{q_d (T)}{T} \approx \Tilde{U_\pm} |\hat t|^{- \alpha} 
\end{equation}
in the vicinity of a CEP, where $\Tilde{U_\pm} =  (1 - \alpha) \,\,  U_\pm(0)$. This divergence is more tractable in experiments as it occurs in the first derivative of $q_d (T)$. 

\begin{figure}
     \centering
         \includegraphics[scale=0.8]{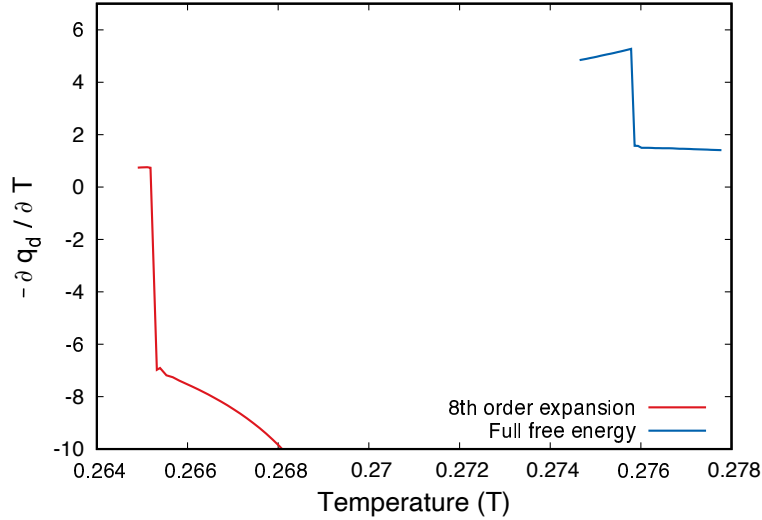}
    \caption{ (colour online) The plot of  the negative phase co-existence diameter derivative ($ -\frac{\partial q_d}{\partial T}$) as a function of $T$ near a CEP for \textbf{(a)} $8^{th}$ order Landau free energy functional and \textbf{(b)} full free energy functional of RCFBC model at $p=0.044$.  The diameter derivative shows jump at the corresponding CEPs. }
      \label{fig76}
 \end{figure}


\subsection{Verification of the singularities in the phase co-existence curve near a CEP }

 \begin{figure}
\centering
     \begin{subfigure}[b]{0.49\textwidth}
         \centering
         \includegraphics[width=\textwidth]{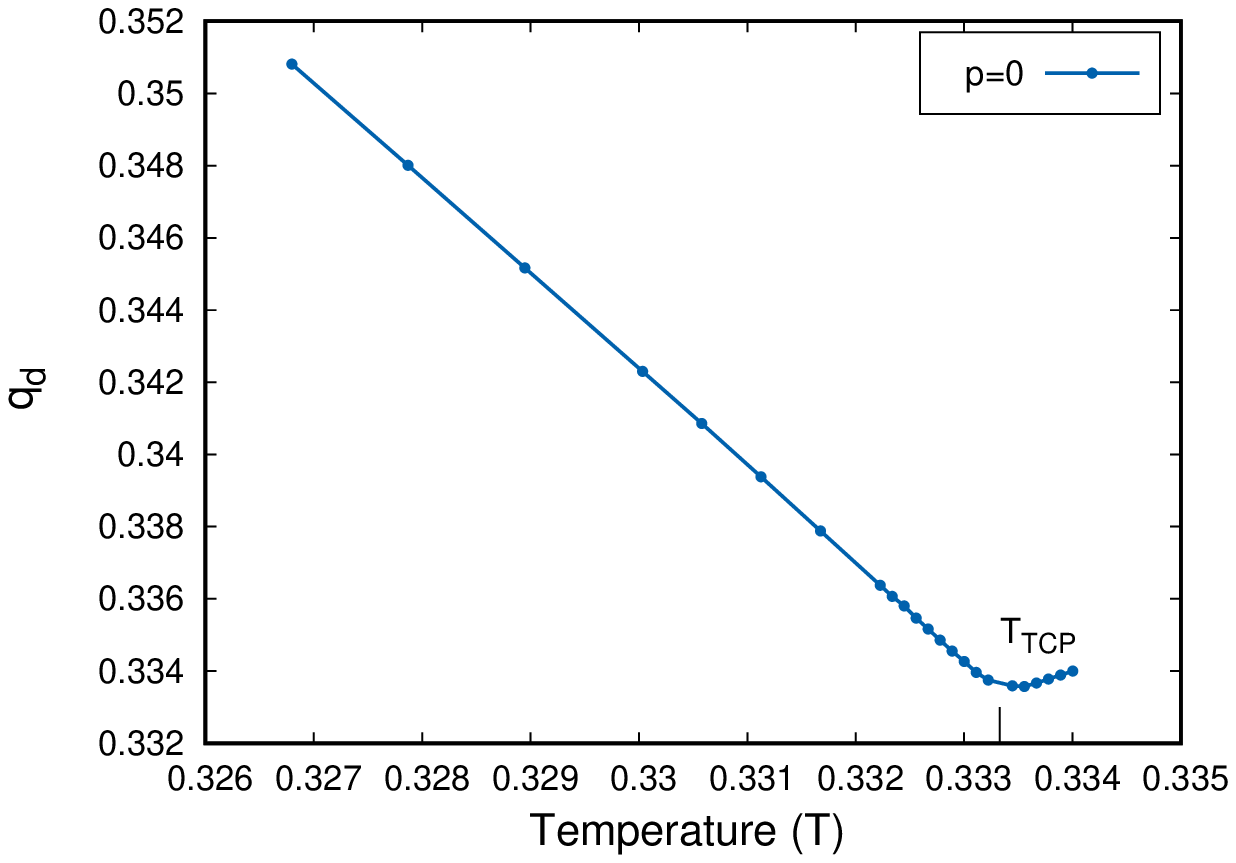}
         \caption{}
         \label{fig580}
     \end{subfigure}
     \hfill
     \centering
     \begin{subfigure}[b]{0.49\textwidth}
         \centering
         \includegraphics[width=\textwidth]{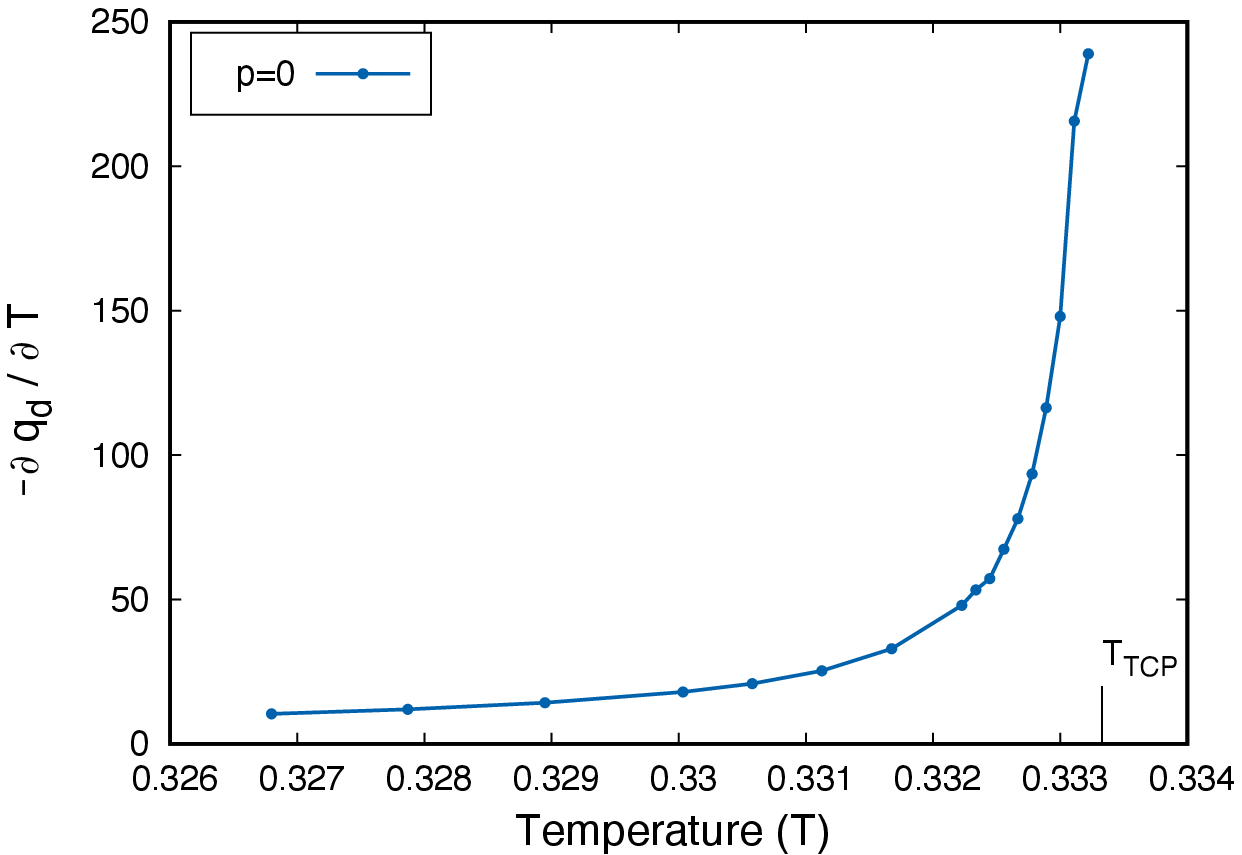}
         \caption{}
         \label{fig581}
     \end{subfigure}
     \hfill
     \begin{subfigure}[b]{0.49\textwidth}
         \centering
        \includegraphics[width=\textwidth]{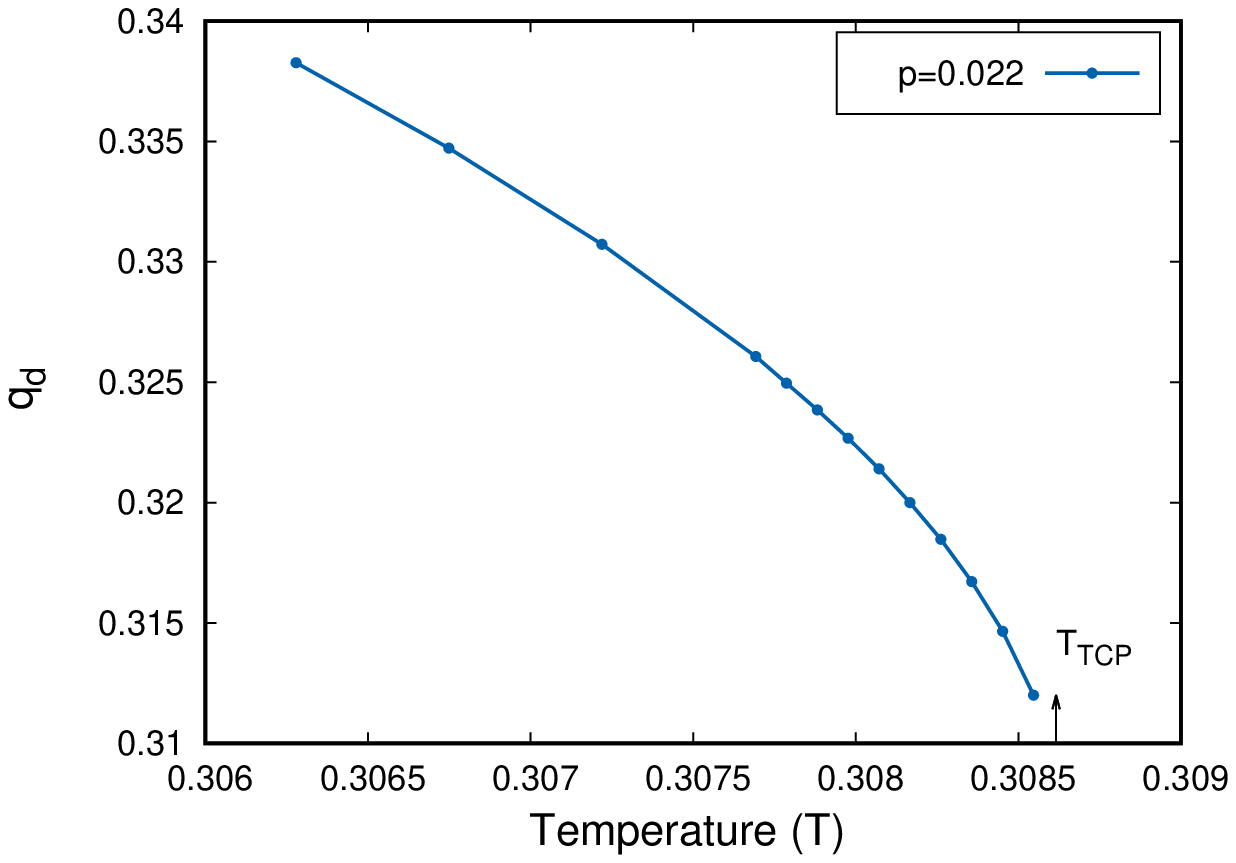}
         \caption{}
         \label{fig582}
     \end{subfigure}
     \hfill
     \centering
     \begin{subfigure}[b]{0.49\textwidth}
         \centering
         \includegraphics[width=\textwidth]{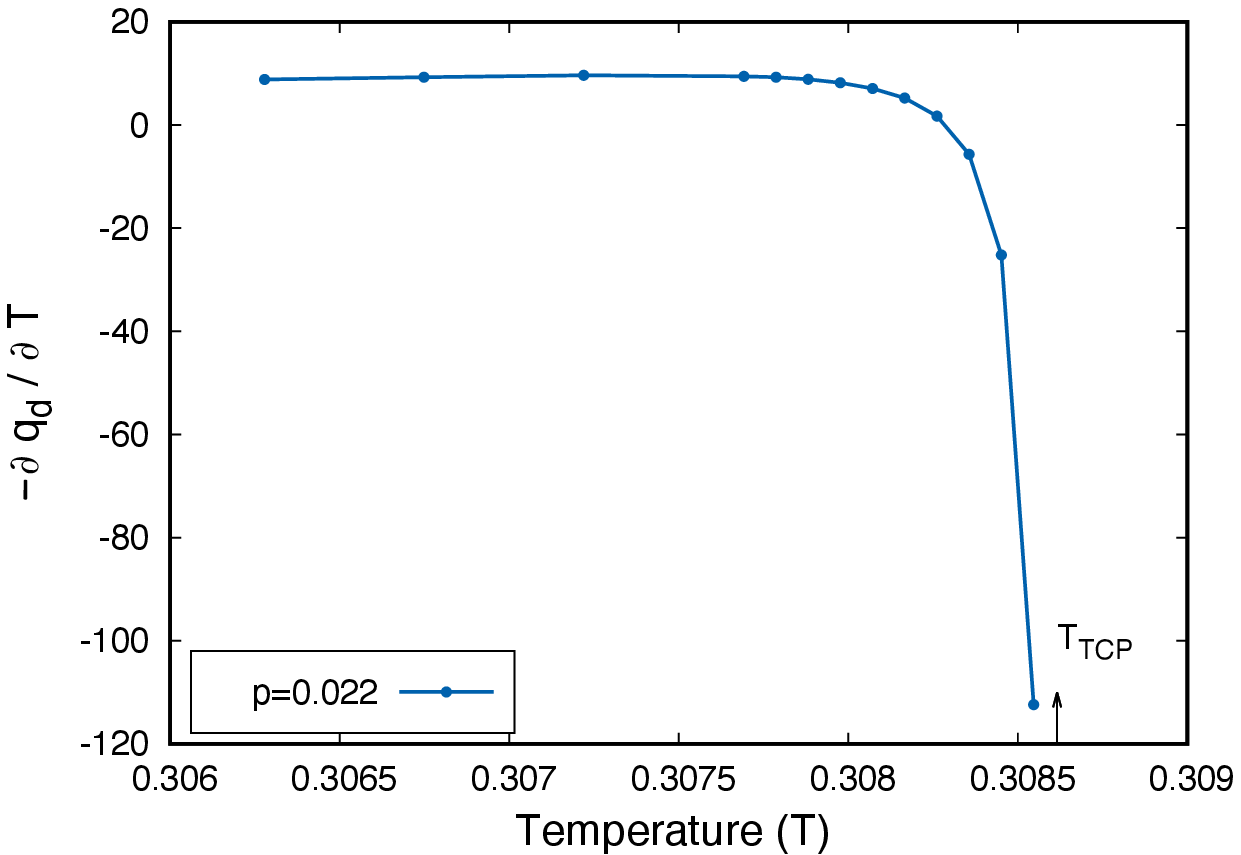}
         \caption{}
         \label{fig583}
     \end{subfigure}
     \hfill
        \caption{The plot of the  co-existence diameter $q_d$, the negative derivative of the diameter $- \frac{\partial q_{q}}{\partial T}$  as a function of $T$ in order to compare the scaling arguments near a TCP for RCFBC model at $p=0$ and $p=0.022$. \textbf{(a)} and \textbf{(c)} show the plot  of $q_d$  as a function of $T$. \textbf{(b)} and \textbf{(d)} show the plot of $- \frac{\partial q_{d}}{\partial T}$ as a function of $T$. The derivative shows a divergence at the TCP.}
        \label{fig58}
\end{figure}

In order to  verify the  Wilding's scaling argument near a CEP, we plot in Fig. \ref{fig57} the co-ordinates of the diameter $q_d$ and the negative of the first derivative $- \frac{\partial q_d}{\partial T}$ of the co-existence curve as a function of $T$. We consider the  RCFBC model for $p=0.044$, $p=0.07$ and RBEG model for $K=-0.6$. In Fig.  \ref{fig570}, \ref{fig572} and \ref{fig602} we plot the diameter $q_d$ of the density co-existence region as a function of $T$. We observe that at the $T_{CEP}$ it shows a kink in $q_d$. In  Fig. \ref{fig571},  \ref{fig573} and  \ref{fig603} we plot  the corresponding derivative $-\frac{\partial q_{d}}{\partial T}$ as a function of $T$. It shows a discontinuity at the CEP. This discontinuity is again similar to the jump of $C_v$ vs $T$ plot with the critical exponent $\alpha=0$.

As mentioned previously in Sec \ref{sec2}, the $8^{th}$ order Landau free energy functional cannot locate the CEP correctly \cite{mukherjee2020emergence}. The singularity in the phase boundary near a CEP has been verified for an $8^{th}$ order Landau free energy functional expansion of an isomorphous transition in \cite{DESANTAHELENA1994479}. In this section we compare the  phase co-existence diameter using both the full free energy functional and by truncating the free energy upto $8^{th}$ order for RCFBC model. For example, Fig. \ref{fig76} shows the plot of the phase co-existence diameter derivative near a CEP predicted by the $8^{th}$ order Landau free energy functional (shown by solid red line) and by the full free energy functional Eq. \ref{eq26} (shown by solid blue line) for $p=0.044$. In both the plots the derivative of the diameter shows a jump near the corresponding values of the CEP. The value of $T_{CEP}$ though is underestimated by the $8^{th}$ order Landau expansion.



Similarly, in order to compare the  Wilding's scaling argument, we plot the diameter of the co-existence region $q_d$ and its derivative $- \frac{\partial q_{d}}{\partial T}$ as a function of $T$ for the RCFBC model at $p=0$ and $p=0.022$ in Fig. \ref{fig580} - \ref{fig583}. We observe that the $q_d$ decreases continuously as a function of $T$, shown in Fig. \ref{fig580} and Fig. \ref{fig582} for $p=0$ and $p=0.022$ respectively. The derivative of the diameter $q_d$ shows a divergence at the TCP as shown in Fig. \ref{fig581} and Fig. \ref{fig583} for $p=0$ and $p=0.022$ respectively.


\section{Conclusion}

Both the TCP and CEP are the multicritical points where a $\lambda$ line meets a first order phase boundary. Experimentally it  gets tricky to distinguish between these two multicritical points by looking at the critical exponents alone.

In this work, we revisit the scaling hypothesis near a CEP for a spin-1 model in the presence of disorder (the RCFBC model) and in the presence of repulsive biquadratic exchange interaction  (the RBEG model). Both these models have TCP and CEP in the phase diagram depending on the strength of disorder and repulsive interaction respectively. We determine the first order phase boundary in the $T-\Delta$ plane near the CEP. We observed that the phase boundary curvature shows a jump at the CEP which confirms the predictions by Fisher \textit{et al} \cite{fisher1990phases, PhysRevB.43.11177}. This jump is similar to the jump in specific heat ($C_v$) as a function of $T$ with the critical exponent $\alpha =0$. On the contrary, we observed that the phase boundary doesn't exhibit any singularity near a TCP.

We also studied the phase co-existence curve near a TCP and CEP for different parameter values in order to observe the change in the shape of the co-existence curve. We observe that the phase co-existence diameter shows a kink near the CEP and it's derivative shows a jump at the CEP as predicted  by Wilding \cite{PhysRevLett.78.1488}. On the other hand near a TCP, the co-existence diameter decays continuously and its derivative diverges.

The above analysis has been done on a fully connected  graph but if the multicritical points like TCP and CEP are present in finite dimensions,  the scaling behaviour  near these multicritical points will be similar.  In finite dimensions also  the phase diagram changes as the disorder strength or the higher order interaction changes.   This is evident from the experiments on   $^3$He-$^4$He mixture in aerogel in \cite{chan1996helium, PhysRevLett.71.2268}.  The qualitative change in the phase diagram with the change in disorder strength is similar to the behaviour of RCFBC model on a fully connected graph \cite{mukherjee2020emergence}.   In the absence of exact results, numerics plays an important role in getting the behaviour of the model in finite dimensions. But locating multicritical points is nontrivial and challenging in numerics  \cite{PhysRevE.92.022134, PhysRevE.105.054143}.  The quantities studied in this work are numerically more tractable and should be useful in classifying the multicritical points in numerical studies of disorder in finite dimensions.


\end{document}